\documentclass[aps,prx,twocolumn,longbibliography]{revtex4-1}
\usepackage{amsmath,amsthm,amsfonts,amssymb}
\usepackage{xcolor,bm}
\usepackage{mathrsfs}
\usepackage{graphicx}
\usepackage{dsfont}
\usepackage{hyperref}
\usepackage{braket}
\usepackage{todonotes}
\usepackage{algorithm, algpseudocode}
\usepackage{algorithmicx}

\newtheorem*{remark*}{Remark}

\begin{document}
\title{Symmetry protected topological order as a requirement for 
measurement-based quantum gate teleportation}

\author{Zhuohao Liu}
\affiliation{Institute for Quantum Science and Technology and Department of
Physics and Astronomy, University of Calgary, Calgary, Alberta T2N 1N4, Canada}
\author{Emma C. Johnson}
\affiliation{Institute for Quantum Science and Technology and Department of
Physics and Astronomy, University of Calgary, Calgary, Alberta T2N 1N4, Canada}
\author{David L. Feder}
\affiliation{Institute for Quantum Science and Technology and Department of
Physics and Astronomy, University of Calgary, Calgary, Alberta T2N 1N4, Canada}

\begin{abstract}

All known resource states for measurement-based quantum teleportation in
correlation space possess symmetry protected topological order, but is this a 
sufficient or even necessary condition? This work considers two families of
one-dimensional qubit states to answer this question in the negative. The first 
is a family of matrix-product states with bond dimension two that includes the 
cluster state as a special case, protected by a global non-onsite symmetry, 
which is characterized by a finite correlation length and a degenerate 
entanglement spectrum in the thermodynamic limit but which is unable to 
deterministically teleport a universal set of single-qubit gates. The second 
are states with bond dimension four that are a resource for deterministic 
universal teleportation of finite single-qubit gates, but which possess no 
symmetry.

\end{abstract}

\maketitle

%\section{Introduction}
%\label{sec:introduction}

% {\it Introduction}---The measurement-based model of quantum computation 

\section{Introduction}
The measurement-based model of quantum computation 
(MBQC)~\cite{Briegel2009,Wei2021} is wholly equivalent to the quantum circuit 
model in its ability to effect arbitrary quantum gates~\cite{Raussendorf2003},
but is advantageous for practical implementations where the application of 
local entangling gates on demand is challenging. In MBQC, the entanglement is 
present at the outset, in the form of a specific resource state, and quantum 
gates are teleported by means of adaptive single-qubit measurements. A
long-standing open problem has been to identify the essential characteristics 
required of a resource state, and accordingly much attention has been focused 
on one-dimensional systems which are able to perform measurement-based gate 
teleportation (MBQT) of arbitrary single-qubit gates. To date, all resource 
states for MBQT that have been identified possess symmetry protected 
topological (SPT) 
order~\cite{Chen2013,Senthil2015,Else2012a,Else2012b,Stephen2017}, which 
passively protects the quantum information from certain kinds of 
errors~\cite{Pollmann2012}; these include the cluster states of the original 
one-way quantum computation model~\cite{Raussendorf2001,Raussendorf2003}, and 
Haldane-phase states~\cite{Haldane1983} such as the ground states of the 
Affleck-Kennedy-Lieb-Tasaki (AKLT) state~\cite{Affleck1987,Miyake2010}, its
generalizations to two dimensions and higher 
spin~\cite{Cai2010,Wei2011,Miyake2011,Wei2013,Wei2015}, and a two-dimensional 
state with genuine SPT order~\cite{Miller2016}.

For all resource states with SPT order, MBQT is performed in correlation space 
in the matrix product state (MPS) 
representation~\cite{Vidal2003,Perez2007,Gross2007,Chen2010}. The group 
cohomology~\cite{Chen2013} ensures that the teleported gate in correlation 
space can be expressed as a rotation operator in a tensor product with an 
unimportant `junk' matrix~\cite{Else2012a,Else2012b}. Unfortunately, the 
teleported gates throughout the universal SPT phase are not strictly in the 
protected `wire basis,' which restricts the target teleported gates to 
infinitesimal rotations in correlation space~\cite{Stephen2017,Raussendorf2019} 
except for the cluster state itself. Another key signature of SPT is the 
degeneracy of the entanglement 
spectrum (ES)~\cite{Pollmann2010,Cirac2011,Else2012b}.

While SPT order has been a powerful approach to classifying resource states for
MBQT, a plethora of key questions remain, even for the simplest case of 
one-dimensional qubits. Are resource states with SPT order required for MBQT? 
If not, what other kinds of resource states are possible? Can any resources
other than the cluster state effect the teleportation of universal single-qubit
gates based on finite, rather than infinitesimal, unitary rotations? What is 
the relationship between the ability of a state to be a resource for MBQT and
the structure of the teleported gates? 

This work partially addresses these questions by considering two specific 
examples. The first is a family of SPT states with bond dimension $D=2$ that 
includes the cluster state as a special case, which is protected by a global 
$\mathbb{Z}_2\times\mathbb{Z}_2$ symmetry that is generally neither unitary nor 
onsite. States within this family have finite correlation length and exhibit 
a degenerate ES for arbitrary boundary conditions in the thermodynamic limit,
but (except for the cluster state itself) are unable to deterministically 
teleport a universal set of protected single-qubit gates in correlation space.
The second example is an extension of the cluster state to a family of non-SPT 
states with bond dimension $D=4$, which are a resource for the deterministic 
teleportation of single-qubit gates, based on finite rotations. The results 
demonstrate that SPT order is neither sufficient nor necessary for a state to 
be an MBQT resource.

\section{Technical background}
\label{sec:tech}

A one-dimensional state for $n$ qubits can be written in the MPS representation 
as
\begin{equation}
|\psi\rangle=\sum_{i_1,\ldots,i_n}A^{[n]}\left[i_n\right]\cdots 
A^{[1]}\left[i_1\right]|i_1\cdots i_n\rangle.
\label{eq:MPS}
\end{equation}
The $A^{[k]}\left[i_k\right]$, where $i_n=\{0,1\}$, are rectangular matrices
and their product (indexed by the strings $i_1\cdots i_n$) therefore constitute 
the amplitudes of $|\psi\rangle$ in the computational basis. Given that 
$A^{[n]}\left[i_n\right]$ ($A^{[1]}\left[i_1\right]$) is a row (column) vector, 
it is conventional to introduce matrices $B^{[k]}[i_k]$ and boundary vectors
$|L\rangle$ and $|R\rangle$ such that $A^{[n]}[i_n]=\langle R|B^{[n]}[i_n]$,
$A^{[1]}[i_1]=B^{[1]}[i_1]|L\rangle$, and $A^{[k]}[i_k]=B^{[k]}[i_k]$ 
otherwise, in which case Eq.~(\ref{eq:MPS}) becomes
\begin{equation}
|\psi\rangle=\sum_{i_1,\ldots,i_n}\langle R|B^{[n]}\left[i_n\right]\cdots 
B^{[1]}\left[i_1\right]|L\rangle|i_1\cdots i_n\rangle.
\label{eq:MPS2}
\end{equation}
In this work, the MPS matrices $B^{[k]}[0]$ and $B^{[k]}[1]$ are arbitrary 
complex matrices with fixed `bond' dimension $D$, and are normalized in (left) 
site canonical form, $\sum_{i_k}{B^{[k]}[i_k]}^{\dag}B^{[k]}[i_k]=I$ for each 
$k$, where $I$ is the identity matrix.

A general single-qubit measurement can be effected by first performing a 
unitary gate $\tilde{U}$ on the qubit, and then measuring it in the 
computational basis by projecting the result onto $|m\rangle\langle m|$, 
$m=0,1$. This is equivalent to applying the operator $|m\rangle\langle m|
\tilde{U}=|m\rangle\langle\phi_m|$, where $|\phi_m\rangle=\tilde{U}^{\dag}
|m\rangle$ constitute a basis for the unitary:
\begin{eqnarray}
|\phi_0\rangle&=&e^{-i\varphi_1}\cos\vartheta|0\rangle
+e^{-i\varphi_2}\sin\vartheta|1\rangle;\nonumber \\
|\phi_1\rangle&=&e^{i\varphi_2}\sin\vartheta|0\rangle
-e^{i\varphi_1}\cos\vartheta|1\rangle.
\label{eq:basis}
\end{eqnarray}
Consider the action of this operator on the first qubit of $|\psi\rangle$ in 
Eq.~(\ref{eq:MPS2}). Ignoring normalization, one obtains
\begin{eqnarray}
&&|m\rangle_1\langle\phi_m|\psi\rangle=|m_1\rangle
\langle R|\sum_{i_2,\ldots,i_n} B^{[n]}\left[i_n\right]\cdots 
B^{[2]}\left[i_2\right]\nonumber \\
&&\qquad\times\left(\sum_{i_1}\langle\phi_m|i_1\rangle B^{[1]}\left[i_1\right]
\right)|L\rangle|i_2\cdots i_n\rangle\nonumber \\
&&\quad=|m\rangle_1\sum_{i_2,\ldots,i_n}\langle R|B^{[n]}\left[i_n\right]
\cdots B^{[2]}\left[i_2\right]|L'\rangle|i_2\cdots i_n\rangle,\hphantom{aa}
\end{eqnarray}
where the left boundary state in correlation space is transformed into
$|L'\rangle=B^{[1]}\left[\phi_m\right]|L\rangle$ by the operator
$B^{[1]}\left[\phi_m\right]=\sum_{i_1}\langle\phi_m|i_1\rangle B^{[1]}[i_1]$;
in general, one obtains
\begin{eqnarray}
B^{[k]}[\phi_0]&=&e^{i\varphi_1}\cos\vartheta B^{[k]}[0]
+e^{i\varphi_2}\sin\vartheta B^{[k]}[1];
\label{eq:outputgate0} \\
B^{[k]}[\phi_1]&=&e^{-i\varphi_2}\sin\vartheta B^{[k]}[0]
-e^{-i\varphi_1}\cos\vartheta B^{[k]}[1],\hphantom{aa}
\label{eq:outputgate1}
\end{eqnarray}
for measurements of 0 and 1, respectively.
Successive measurements therefore apply a sequence of gates to $|L\rangle$.
For MBQT, however, $B^{[k]}\left[\phi_m\right]$ must correspond to a unitary 
operator for all $m$, a severe restriction on possible resource states which 
are defined by the matrices $B^{[i_k]}[i_k]$.

One-dimensional cluster states of qubits with open boundary conditions provide 
a convenient reference for the work presented here. It is straightforward to 
verify that these are (non-uniquely) described by an MPS representation with 
matrices
\begin{eqnarray}
B^{[k]}[0]&=&\frac{1}{\sqrt{2}}
\begin{pmatrix} 
1 & 0\cr 1 & 0\cr
\end{pmatrix}=|+\rangle\langle 0|;\nonumber \\
B^{[k]}[1]&=&\frac{1}{\sqrt{2}}
\begin{pmatrix} 
0 & 1\cr 0 & -1\cr
\end{pmatrix}=|-\rangle\langle 1|
\label{eq:MPS4cluster}
\end{eqnarray}
for $k=1,\ldots,n$, where 
$|\pm\rangle=\left(|0\rangle\pm|1\rangle\right)/\sqrt{2}$, 
$\langle R|=\langle 0|$, and $|L\rangle=\sqrt{2}|+\rangle$.
Because the matrices are independent of site, the MPS representation is said to 
be translationally invariant, even though the state itself is defined with open 
boundary conditions. One obtains
\begin{equation}
B^{[k]}[\phi_m]=\langle\phi_m|0\rangle|+\rangle\langle 0|
+\langle\phi_m|1\rangle|-\rangle\langle 1|,
\end{equation}
which is unitary if
\begin{eqnarray}
|\langle\phi_m|0\rangle|^2|+\rangle\langle +|
+|\langle\phi_m|1\rangle|^2|-\rangle\langle -|&=&cI;\nonumber \\
|\langle\phi_m|0\rangle|^2|0\rangle\langle 0|
+|\langle\phi_m|1\rangle|^2|1\rangle\langle 1|&=&cI,
\end{eqnarray}
where $c$ is a constant related to the (re)normalization of $|\psi\rangle$ 
after the measurement. These two conditions require $|\langle\phi_m|0\rangle|^2
=|\langle\phi_m|1\rangle|^2=c$ for $m=0,1$, and it is straightforward to verify 
that they together imply $B^{[k]}[\phi_m]=X^mHR_Z(\theta)$ and $c=1/2$ 
ignoring overall phase factors, where 
\begin{equation}
X=\begin{pmatrix}
0 & 1\cr
1 & 0
\end{pmatrix};\;
Z=\begin{pmatrix}
1 & 0\cr
0 & -1
\end{pmatrix};\;
H=\frac{1}{\sqrt{2}}\begin{pmatrix}
1 & 1\cr
1 & -1
\end{pmatrix},
\end{equation}
and $R_Z(\theta)=\exp(iZ\theta)$. 

\section{SPT states unable to effect MBQT}
\label{sec:SPT}

\subsection{MPS matrices and the state}
\label{subsec:MPSandstate}

The (unitary) teleported gate~(\ref{eq:outputgate0}) is chosen to have the form
$yU$, where 
\begin{equation}
U=\begin{pmatrix}
e^{i\phi_1}\cos\theta & e^{i\phi_2}\sin\theta\cr
e^{-i\phi_2}\sin\theta & -e^{-i\phi_1}\cos\theta\cr
\end{pmatrix}
\label{eq:U}
\end{equation}
with all parameters assumed to be real, and $y$ is a proportionality factor to 
account for the renormalization of the state after measurement; this case 
corresponds to an MPS with bond dimension $D=2$. Consider first the simplest 
case of a translationally invariant system. The derivation is straightforward 
but unwieldy, and is relegated to Appendix~\ref{app:MPSderivation}. Choosing
the MPS matrices to be in column form as in the cluster-state case, 
Eq.~(\ref{eq:MPS4cluster}), and ensuring that they do not depend on the 
measurement angles, restricts both the measurement basis and the teleported
gates; one choice corresponds to $\varphi_2=-\varphi_1:= -\varphi$,
$\vartheta=(2k+1)\pi/4$, $k\in\mathbb{Z}$, and $y=1/\sqrt{2}$. These yield 
the measurement basis $\tilde{U}^{\dag}=HR_Z(\varphi)$, exactly as in 
cluster-state teleportation. One obtains the (non-unique) expressions for the 
MPS matrices, Eq.~(\ref{eq:MPS4new}):
\begin{equation}
B[0]=\begin{pmatrix}
\cos\theta & 0\cr
\sin\theta & 0\cr
\end{pmatrix};\;
B[1]=\begin{pmatrix}
0 & \sin\theta\cr
0 & -\cos\theta\cr
\end{pmatrix},
\label{eq:BUtensorJ}
\end{equation}
which in turn yield the measurement-dependent teleported unitary gates, 
Eq.~(\ref{eq:U4new}):
\begin{eqnarray}
U[0]&=&ZR_Y(\theta)R_Z(\varphi);
\label{eq:U0} \\
U[1]&=&R_Y(-\theta)R_Z(\varphi)=ZR_Y(\theta)ZR_Z(\varphi)=U_0Z,\hphantom{aa}
\label{eq:U1}
\end{eqnarray}
where $Y=iXZ$. 

Assuming a translationally invariant MPS, consistently measuring $|0\rangle$ 
would yield successive rotations about $Z$ and 
$R_Y(-\theta)R_Z(\varphi)R_Y(\theta)$, a measurement-dependent rotation around 
$Z$ conjugated by a fixed rotation around $Y$. These are non-parallel axes,
which allows for the teleportation of any single-qubit unitary; for the cluster 
state, $\theta=\theta_c:=(2k+1)\pi/4$, $k\in\mathbb{Z}$, and the latter 
rotation axis is $X$. However, the byproduct operator when measuring 
$|1\rangle$ is not easily compensated for. Consider the teleported gates on two 
successive measurements:
\begin{eqnarray}
U^{(2)}U^{(1)}&=&ZR_Y(\theta)Z^{m_2}R_Z(\varphi_2)
ZR_Y(\theta)Z^{m_1}R_Z(\varphi_1)\nonumber \\
&=&Z^{m_1+m_2}R_Y[(-1)^{m_1+m_2+1}\theta]R_Z(\varphi_2)\nonumber \\
&\times&R_Y[(-1)^{m_1}\theta]R_Z(\varphi_1).
\end{eqnarray}
If $m_1=1$, the second rotation corresponds instead to 
$R_Y[(-1)^{m_2}\theta]R_Z(\varphi_2)R_Y(-\theta)$, which if $m_2=0$ is a 
rotation around $Z$ conjugated by a rotation around $Y$ in the opposite 
direction than would be the case for $m_1=0$. One strategy would be to choose 
$\varphi_2=0$ on the next measurement, but if one instead obtains $m_2=1$ one
is left with a second unwanted rotation $R_Y(-2\theta)$ that would somehow need
to be compensated for on the third measurement. The MBQT protocol would 
therefore become non-deterministic. Another strategy might be to restrict the 
$\varphi_i$ to infinitesimal angles, but in this case the error induced by the 
byproduct is
$R_Y(-\theta)R_Z(\varphi)R_Y(\theta)-R_Y(\theta)R_Z(\varphi)R_Y(-\theta)
\approx i\sin(\theta)\varphi X\to 0$ as $\varphi\to 0$. Because the error
accumulates, the teleported state would be indistinguishable from noise after
several iterations. 

The inability of a state defined by the MPS~(\ref{eq:BUtensorJ}) to effect 
deterministic MBQT of either finite or infinitesimal gates is in marked
contrast from the `oblivious wire' protocol that ensures that SPT states have
uniform computational power to effect 
MBQT~\cite{Raussendorf2017,Raussendorf2019}. In that case the gates are 
infinitesimally displaced from the symmetry-protected wire basis, in order to 
compensate for the fact that the junk matrices are generally 
measurement-dependent. But this is not possible for the simple $D=2$ case under
consideration here. Over all MPS matrices~(\ref{eq:BUtensorJ}), only the 
cluster state can effect deterministic teleportation.

The (unnormalized) state $|\psi\rangle$ can be constructed either directly from 
Eq.~(\ref{eq:MPS2}) or using the machinery in Ref.~\cite{Perez2007}. With left 
and right boundary states 
\begin{equation}
\langle R|=a_R\langle 0|+b_R\langle 1|;\;
|L\rangle=a_L|0\rangle+b_L|1\rangle
\label{eq:tensorproducts}
\end{equation}
and site-dependent 
values of $\theta$, the state takes an especially simple form after some 
straightforward algebra:
\begin{eqnarray}
|\psi\rangle&=&\prod_{j=1}^{n-1}C_{\theta_j}^{(j,j+1)}
\left(x_1|0\rangle+x_2|1\rangle\right)\otimes|+\rangle^{\otimes n-2}\nonumber \\
&\otimes&\left(a_L|0\rangle+b_L|1\rangle\right),
\label{eq:tensorstate1}
\end{eqnarray}
$x_1=a_R\cos\theta_n+b_R\sin\theta_n$, $x_2=a_R\sin\theta_n-b_R\cos\theta_n$, 
and
\begin{equation}
C_{\theta_j}^{(j,j+1)}=\sqrt{2}{\rm diag}(\cos\theta_j,\sin\theta_j,
\sin\theta_j,-\cos\theta_j)_{j,j+1}
\label{eq:Ctheta}
\end{equation}
acts on qubits $j$ and $j+1$. With $\theta_j=\pi/4$ for $j>1$, so that 
$C_{\theta_j}=C_Z=\mbox{diag}(1,1,1,-1)$, the state coincides with the
one-dimensional cluster state with rotated left and right physical qubits.
As $C_{\theta_j}^{j,j+1}$ is only a unitary operator for 
$\theta_j=\theta_c$ $\forall j$, Eq.~(\ref{eq:tensorstate1}) should 
be considered as an expression of the state rather than as a procedure for 
generating it. 

The static correlation function $C_{{\mathcal O}_k,{\mathcal O}'_r}
=\langle{\mathcal O}_k{\mathcal O}_r\rangle
-\langle{\mathcal O}_k\rangle\langle{\mathcal O}_r\rangle$ with respect to
operators ${\mathcal O}_k$ and ${\mathcal O}_r$ acting on sites $k$ and $r$
generically decays exponentially for a 1D MPS with finite bond 
dimension~\cite{Zauner2015} (consistent with the parent Hamiltonian being
gapped~\cite{Fannes1992,Nachtergaele1996,Hastings2004,Brandao2015}):
$\left\Vert C_{{\mathcal O}_k,{\mathcal O}_r'}\right\Vert\sim e^{-|k-r|/\xi}$,
where $\xi$ is the correlation length. For a translationally invariant MPS, 
$\xi=-1/\ln(\lambda_1)$, where $\lambda_1$ is the second-largest eigenvalue of
the transfer matrix~\cite{Verstraete2008}
\begin{equation}
T_k=\sum_{i_k}{B^{[k]}[i_k]}^*\otimes B^{[k]}[i_k].
\end{equation}
Using Eq.~(\ref{eq:BUtensorJ}), one obtains $\lambda_0=1$ and 
$\lambda_1=\cos 2\theta$; thus, $\xi=-1/\ln(\cos 2\theta)$. The correlation
length is zero when $\theta=\theta_c$, it diverges as $\xi\sim 1/2\theta^2$ for 
$\theta\to 0$, 

\subsection{SPT order}
\label{subsec:SPTorder}

The real-space representation of the state, Eq.~(\ref{eq:tensorstate1}), allows
for the explicit construction of the symmetry operators. As shown in 
Appendix~\ref{app:SPTorder}, the state possesses an exact 
$\mathbb{Z}_2\times\mathbb{Z}_2$ symmetry $O(g_1,g_2)|\psi\rangle=|\psi\rangle$,
where $O(g_1,g_2)=X^{g_1}_{\rm odd}X^{g_2}_{\rm even\vphantom{d}}$ and 
$g_1,g_2\in\{0,1\}$. The operators $X_{\rm odd}$ and $X_{\rm even}$ are the 
analogs of the $X$ symmetry operators that act on odd-labeled and even-labeled 
sites of the cluster state, respectively, and for an even number of sites are 
given by Eq.~(\ref{eq:symmetryops}):
\begin{eqnarray}
X_{\rm odd}&=&P_{\theta_1}^{1,2}\left(X_1\prod_{j=1}
P_{\theta_{2j}}^{2j,2j+1}P_{\theta_{2j+1}}^{2j+1,2j+2}X_{2j+1}\right)Z_n;
\nonumber \\
X_{\rm even}&=&Z_1\left(\prod_jP_{\theta_{2j-1}}^{2j-1,2j}
P_{\theta_{2j}}^{2j,2j+1}X_{2j}\right)\left(P_{\theta_n}^{n-1,n}X_n\right),
\nonumber \\
\label{eq:symmetryops2}
\end{eqnarray}
where
\begin{equation}
P_{\theta_j}^{j,j+1}=\mbox{diag}(\cot\theta_j,\tan\theta_j,\tan\theta_j,
\cot\theta_j)_{j,j+1}.
\end{equation}
Alternatively, these can be written as 
$X_{\rm odd}=S_{1,2}\left(\prod_jS_{2j+1,2j+2}\right)$ and
$X_{\rm even}=\left(\prod_jS_{2j,2j+1}\right)S_{n-1,n}$, where
$S_{j,j+1}=P_{\theta_j}^{j-1,j}P_{\theta_{j+1}}^{j,j+1}Z_{j-1}X_jZ_{j+1}$ are
(nonlocal) stabilizer generators for the state, Eq.~(\ref{eq:tensorstate1}). 
While these symmetry operators square to the identity and commute with one 
another, as shown in Appendix~\ref{app:SPTorder}, they are neither unitary nor 
onsite. 

Consider the left boundary qubit. The $X$ and $Z$ gates are transformed by the 
$C_{\theta}$ operators into effective Pauli gates: 
$\overline{X}_1=P_{\theta_1}^{1,2}X_1Z_2$ and $\overline{Z}_1=Z_1$. Then one 
may determine the effective operators $\overline{X}_1'$ and $\overline{Z}_1'$ 
corresponding to $\overline{X}_1$ and $\overline{Z}_1$ conjugated by 
$O(g_1,g_2)$, respectively. Straightforward algebra presented in 
Appendix~\ref{app:SPTorder} reveals $\overline{Z}_1'=(-1)^{g_1}\overline{Z}_1$
and $\overline{X}_1'=(-1)^{g_2}\overline{X}$. The transformations on 
$\overline{Z}$ and $\overline{X}$ by the $\mathbb{Z}_2\times\mathbb{Z}_2$
operators are therefore equivalent to conjugation under an effective operator
$O_{\rm eff}(g_1,g_2)=\overline{X}^{g_1}\overline{Z}^{g_2}$, which is the same 
as for the regular cluster state. A similar result holds for the right 
boundary. Thus, the state belongs to the same maximally non-commutative 
phase as the cluster state~\cite{Else2012a,Stephen2017}.

In the cluster-state limit $\theta_j=\theta_c\;\forall j$, the symmetry 
operators~(\ref{eq:symmetryops2}) reduce to 
$X_{\rm odd}=\prod_{j=1}X_{2j-1}$ and $X_{\rm even}=\prod_{j=1}X_{2j}$, as 
expected. The onsite symmetry $U(g)=U(g_1,g_2)
=X_{\rm odd}^{g_1}X_{\rm even}^{g_2}$, which acts in parallel on adjacent 
two-site blocks so that $U(g)^{\otimes n/2}|\psi\rangle=|\psi\rangle$, is 
shared by the MPS matrices themselves via~\cite{Perez2007,Perez2008,Schuch2011}
\begin{equation}
\sum_{\bf j}U(g)_{{\bf i},{\bf j}}A[{\bf j}]=e^{i\phi_g}V(g)A[{\bf i}]
V(g)^{\dag},
\label{eq:injectivity}
\end{equation}
where ${\bf i}$ and ${\bf j}$ are bitstrings of length 2, and
$A[{\bf i}]=A[i_1i_2]:=A[i_1]\otimes A[i_2]$. It is straightforward to verify
that $V(g)=Y_1^{g_1}\otimes Y_2^{g_2}$ and $\phi_g=0$. For any choices of 
$\theta_j\neq\theta_c$, however, $O(g)=O(g_1,g_2)$ is non-onsite, and there is 
no analog of Eq.~(\ref{eq:injectivity}) that can be expressed in 
block-injective form for any length smaller than $n$. Rather,
\begin{equation}
\sum_{\bf j}O(g)_{{\bf i},{\bf j}}A[{\bf j}]
\neq e^{i\phi_g}V({\bf g})A[{\bf i}]V({\bf g})^{-1},
\label{eq:injectivity2}
\end{equation}
where ${\bf i}$, and ${\bf j}$ are now bitstrings of length $n$ and
${\bf g}=(1010\cdots)^{g_1}\oplus(0101\cdots)^{g_2}$ ($\oplus$ is bitwise 
addition mod 2), for any $V({\bf g})$: the only non-zero term on the left is
${\bf j}={\bf i}\oplus{\bf g}$, and $|O(g)_{{\bf i},{\bf i}\oplus{\bf g}}|
\neq 1$. Thus, the symmetry of the real-space state is no longer shared by the 
(product of) MPS matrices for a global non-onsite symmetry. 

\subsection{Entanglement Spectrum}
\label{subsec:entanglement}

Consistent with the SPT order of the state defined either by the MPS 
matrices~(\ref{eq:MPS4cluster}) or the state~(\ref{eq:tensorstate1}), the ES
is asymptotically degenerate in the thermodynamic limit for all choices of 
boundary conditions. The ES corresponds to the eigenvalues of the reduced 
density matrix associated with a partition of the one-dimensional state with 
$\ell$ qubits on the left 
and $n-\ell$ qubits on the right. It can be obtained by diagonalizing the 
reduced density matrix, but more efficiently from the MPS matrices. Following 
Prosen~\cite{Prosen2006}, one may express the amplitudes of the 
state~(\ref{eq:MPS2}) as
\begin{equation}
\langle R|B^{[n]}\left[i_n\right]\cdots
B^{[1]}\left[i_1\right]|L\rangle=\sum_{j=1}^D\Phi_{\ell,j}^R\Phi_{\ell,j}^L;
\hphantom{aaa}
\end{equation}
here,
\begin{equation}
\Phi_{\ell,j}^R:=\langle R|B^{[n]}_n\cdots B^{[\ell+1]}_{\ell+1}|j\rangle;\;
\Phi_{\ell,j}^L:=\langle j|B^{[\ell]}_{\ell}\cdots B^{[1]}_1|L\rangle,
\label{eq:Phi}
\end{equation}
and $|j\rangle$ are computational basis states. 
The elements of covariance matrices $V_n^L$ and $V_n^R$ are obtained via
\begin{eqnarray}
\langle j'|V_{\ell}^R|j\rangle&:=&\sum_{\ell+1,\ldots,n}\Phi_{\ell,j'}^{R*}
\Phi_{\ell,j}^R;\nonumber \\
\langle j'|V_{\ell}^L|j\rangle&:=&\sum_{1,\ldots,\ell}
\Phi_{\ell,j'}^L\Phi_{\ell,j}^{L*},
\label{eq:V}
\end{eqnarray}
where the sum is over all internal indices. The ES coincides with the 
eigenvalues of $V_{\ell}^RV_{\ell}^L$.

The calculations for the solution~(\ref{eq:BUtensorJ}) with boundary conditions
specified in Eq.~(\ref{eq:tensorproducts}) are given in 
Appendix~\ref{app:entanglement}. For a bulk bipartition where $2<\ell<n-1$, one 
obtains Eq.~(\ref{eq:VLVR}:
\begin{eqnarray}
V_{\ell}^{R}&=&\frac{1}{2}\begin{pmatrix}
1+\alpha & 0\cr
0 & 1-\alpha\cr
\end{pmatrix};\nonumber \\
V_{\ell}^{L}&=&\frac{1}{2}\begin{pmatrix}
1+\beta\cos(2\theta_{\ell}) & \beta\sin(2\theta_{\ell})\cr
\beta\sin(2\theta_{\ell}) & 1-\beta\cos(2\theta_{\ell})\cr
\end{pmatrix},
\end{eqnarray}
where 
$\alpha:=\left(|x_1|^2-|x_2|^2\right)\prod_{k=\ell+1}^{n-1}
\cos(2\theta_k)$ and $\beta:=\left(|a_L|^2-|b_L|^2\right)\prod_{k=1}^{\ell-1}
\cos(2\theta_k)$, with $x_1$, $x_2$ defined below 
Eq.(\ref{eq:tensorstate1}). Note that the matrix elements depend explicitly on 
the boundary states. If $\prod_{k=1}^{\ell-1}\cos(2\theta_k)
=\prod_{k=\ell+1}^{n-1}\cos(2\theta_k)=0$, then $V_{\ell}^{R}$ and 
$V_{\ell}^{L}$ are proportional to the identity and the ES is degenerate. 
This condition is automatically satisfied for the cluster state, 
$\theta_k=\theta_c,\forall k$. If $\theta_k\neq\theta_c$, however, both
matrices are strictly diagonal only if $\alpha$ and $\beta$ do not depend on 
the choice of $\ell$, corresponding to $|x_1|^2=|x_2|^2$ and $|a_L|^2=|b_L|^2$, 
which includes the state that is fully invariant under $O(g_1,g_2)$. In 
general, the (unnormalized) eigenvalues of $V_{\ell}^{R}V_{\ell}^{L}$ are given 
by
\begin{eqnarray}
\lambda_{\pm}&=&\frac{1}{4}\Big\{1+\alpha\beta\cos(2\theta_{\ell})\nonumber \\
&\pm&\sqrt{[1+\alpha\beta\cos(2\theta_{\ell})]^2-(1-\alpha^2)(1-\beta^2)}
\Big\}.
\end{eqnarray}

The ES becomes asymptotically degenerate in the thermodynamic limit,
for any boundary conditions. For $0<\theta_k<\theta_c$, one has 
$0<\cos(2\theta_k)<1$ so that $\alpha,\beta\to 0$ as $n\to\infty$ for any bulk 
bipartition, $\ell\sim n/2$; in that case, $V_{\ell}^L,V_{\ell}^R\to I/2$. 
Alternatively, in the translationally invariant case $\theta_k=\theta$ one may 
write $\cos(2\theta)=e^{-1/\xi}$, where $\xi$ is the correlation length. This 
yields $\alpha=(|x_1|^2-|x_2|^2)e^{-(n-\ell-2)/\xi}$ and
$\beta=(|a_L|^2-|b_L|^2)e^{-(\ell-2)/\xi}$, so that $\alpha,\beta\to 0$ 
exponentially quickly on finite chains as long as $\ell,n-\ell\gg\xi$. In this
aspect, the system behaves much like the AKLT chain~\cite{Geraedts2010}.

To summarize the results of this section: SPT order on qubits is not a 
sufficient condition for the state to be a resource for deterministic MBQT with 
finite or infinitesimal gates. 

\section{Non-SPT states that effect MBQT}

Consider next the case where the teleported gate in the $D=4$ correlation space 
is a direct sum $U\oplus J$ of a $2\times 2$ unitary $U$, given again by 
Eq.~(\ref{eq:U}), and an arbitrary junk matrix
\begin{equation}
J=\begin{pmatrix}
pe^{i\phi_p} & qe^{i\phi_q}\cr
re^{i\phi_r} & se^{i\phi_s}\cr
\end{pmatrix},
\end{equation}
with all parameters real. The $U$ at each measurement step acts on the 
$\{|00\rangle,|01\rangle\}$ computational subspace of the virtual two-qubit 
state, which can be considered as encoding a single qubit, while $J$ acts on 
the complementary subspace. Assuming a direct sum is notationally convenient in 
what follows, but choosing any other subset of registers yields an equivalent 
description. For example, if $U$ and $J$ act on the odd-parity and even-parity 
subspaces $\{|01\rangle,|10\rangle\}$ and $\{|00\rangle,|11\rangle\}$ 
respectively, the output has the characteristic structure of a 
matchgate~\cite{Valiant2002,Brod2011}, and indeed would correspond exactly to a 
matchgate if $\det(U)=\det(J)$.
%If it were possible to control $J$, then the 
%direct-sum or matchgate structure allows for the teleportation of an arbitrary 
%two-qubit gate~\cite{Barenco1995,Divincenzo1995}.

The procedure follows closely the strategy above. Setting 
$\varphi_2=-\varphi_1:=-\varphi$, $\phi_1=\phi_p=\phi_r=\varphi$, and 
$\phi_2=\phi_q=\phi_s=-\varphi$, one obtains
\begin{eqnarray}
B^{[k]}[0]&=&y\sec\vartheta\begin{pmatrix}
\cos\theta & 0 & 0 & 0\cr
\sin\theta & 0 & 0 & 0\cr
0 & 0 & p & 0\cr
0 & 0 & r & 0\cr
\end{pmatrix};\nonumber \\
B^{[k]}[1]&=&y\csc\vartheta\begin{pmatrix}
0 & \sin\theta & 0 & 0\cr
0 & -\cos\theta & 0 & 0\cr
0 & 0 & 0 & q\cr
0 & 0 & 0 & s\cr
\end{pmatrix}.
\label{eq:BUplusJ}
\end{eqnarray}
Enforcing the canonical normalization conditions requires
$\sec^2\vartheta=\csc^2\vartheta$, which is satisfied again by setting
$\vartheta=(2k+1)\pi/4$, $k\in\mathbb{Z}$, in which case the teleported 
single-qubit unitaries coincide 
with Eqs.~(\ref{eq:U0}) and (\ref{eq:U1}) according to the measurement outcome. 
The normalization conditions also require $y=1/\sqrt{2}$ and 
$p^2+r^2=q^2+s^2=1$. These last conditions can be conveniently incorporated by 
setting $p=\cos\gamma$, $r=\sin\gamma$, $q=\sin\delta$, and $s=\cos\delta$, in 
which case the junk matrices become
\begin{equation}
J[0]=\begin{pmatrix}
e^{i\varphi}\cos\gamma & e^{-i\varphi}\sin\delta \cr
e^{i\varphi}\sin\gamma & e^{-i\varphi}\cos\delta \cr
\end{pmatrix};\;J[1]=J[0]Z.
\label{eq:Jsum}
\end{equation}

A major motivation is to explore the possibility that the direct-sum format can
yield new resource states for deterministic MBQT with finite rotations, so the
primary focus is on the $\theta=\pi/4$ case, in which case 
$U[m]=X^mHR_Z(\varphi)$. Assuming site-dependent junk matrices, and general 
boundary states $\langle R|=\{a_R, b_R, c_R, d_R\}$ and
$|L\rangle=\{a_L, b_L, c_L, d_L\}^T$, one obtains the unnormalized states
\begin{equation}
|\psi\rangle=\prod_{j=1}^{n-1}C_Z^{j,j+1}|\psi_1\rangle
+\prod_{j=1}^{n-1}C_{\gamma_j,\delta_j}^{j,j+1}|\psi_2\rangle,
\label{eq:sumstate}
\end{equation}
where
\begin{eqnarray}
|\psi_1\rangle&=&\left(x_1|0\rangle+x_2|1\rangle\right)\otimes|+\rangle^{n-2}
\otimes\left(a_L|0\rangle+b_L|1\rangle\right);\nonumber \\
|\psi_2\rangle&=&\left(x_3|0\rangle+x_4|1\rangle\right)\otimes|+\rangle^{n-2}
\otimes\left(c_L|0\rangle+d_L|1\rangle\right),\hphantom{aaa}
\end{eqnarray}
$x_1=a_R+b_R$, $x_2=a_R-b_R$,
\begin{eqnarray}
x_3&=&\sqrt{2}\left(c_R\cos\gamma_n+d_R\sin\gamma_n\right);\nonumber \\
x_4&=&\sqrt{2}\left(c_R\sin\delta_n+d_R\cos\delta_n\right),
\end{eqnarray}
and 
\begin{equation}
C_{\gamma_j,\delta_j}^{j,j+1}=\sqrt{2}{\rm diag}(\cos\gamma_j,\sin\delta_j,
\sin\gamma_j,\cos\delta_j)_{j,j+1}.
\end{equation}
The state in Eq.~(\ref{eq:sumstate}) allows for the 
teleportation of deterministic single-qubit unitaries (with feed-forward) for 
all choices of junk-matrix angles $\gamma_j$ and $\delta_j$, because the 
computational subspace acts like a cluster state and is orthogonal to (and
therefore remains independent of) the junk subspace.

The matrices~(\ref{eq:BUplusJ}) are in block-diagonal form, so that the MPS is 
not injective~\cite{Perez2007}. This further implies that the 
state~(\ref{eq:sumstate}) cannot be the unique ground state of a local 
frustration-free parent Hamiltonian, but rather that the ground-state 
degeneracy of such a parent Hamiltonian is two, corresponding to the number of 
blocks; this however doesn't preclude the possibility of preparing the state 
directly via a quantum circuit. In principle, the non-injectivity could affect 
readout of the final state~\cite{Stephen2017}. In practice, the state can be 
chosen such that $C_{\gamma_{n-1},\delta_{n-1}}^{n-1,n}=I$ via 
$\gamma_{n-1}=\delta_{n-1}=\pi/4$ so that no entanglement is generated between 
the last two qubits in the junk sector. This prevents any information reaching 
the junk output state $c_L|0\rangle+d_L|1\rangle$, which can be defined in any 
convenient way, and therefore the quantum information encoded in the cluster
sector remains uncontaminated.

Similar to state~(\ref{eq:tensorstate1}), the $C_{\gamma_j,\delta_j}^{j,j+1}$ 
in Eq.~(\ref{eq:sumstate}) are not generally unitary and are (potentially) 
site-dependent. Because~(\ref{eq:sumstate}) is described by a superposition of 
states, each defined by a different set of generalized stabilizers, it no
longer possesses SPT order. Thus, neither SPT order nor injectivity are 
necessary conditions for a state to be a resource for MBQT.

%\section{Conclusions}
%\label{sec:conclusions}

%{\it Conclusions}---The results presented in this work demonstrate that the 
\section{Conclusions and Discussion}

The results presented in this work demonstrate that the presence of 
symmetry-protected topological order is neither a sufficient nor necessary 
condition for a quantum state to be a resource for deterministic 
measurement-based quantum gate teleportation. On the one hand, a family of 
states of one-dimensional qubits with a non-onsite SPT symmetry is unable to 
deterministically teleport universal one-qubit gates in correlation space,
while on the other a family of states with no SPT order is able to do so. All 
identified states can be considered to be analogs of cluster states, but where 
the $C_Z$ entangling gates in their description are generally replaced by 
diagonal non-unitary operators. 

The family of states with non-onsite symmetries identified here belong to the 
same SPT phase as the cluster state, and therefore can be prepared from the 
cluster state via a constant-depth quantum circuit comprised of non-overlapping 
$k$-local unitaries~\cite{Chen2010b}. The fact that such a unitary 
transformation maps a resource state for deterministic MBQT to a non-resource 
state suggests that a large number of states in a given SPT phase may not be
resources for MBQT. Rather, perhaps only the subset of transformations that 
preserve the onsite nature of the symmetry would ensure that the state remains 
within the same computational phase. 

\acknowledgments{The authors are grateful to Tzu-Chieh Wei and Nicholas O'Dea
for helpful comments and suggestions. This work was supported by the Natural 
Sciences and Engineering Research Council of Canada.}

%\pagebreak
%\newpage

\appendix
\begin{widetext}
\vspace{0.2in}

\section{Derivation of MPS matrices}
\label{app:MPSderivation}

Express the MPS matrices as follows
\begin{equation}
B[0]=\begin{pmatrix}
a_{11}e^{i\phi_{11}} & a_{12}e^{i\phi_{12}} \cr
a_{13}e^{i\phi_{13}} & a_{14}e^{i\phi_{14}} \cr
\end{pmatrix};\quad
B[1]=\begin{pmatrix}
a_{21}e^{i\phi_{21}} & a_{22}e^{i\phi_{22}} \cr
a_{23}e^{i\phi_{23}} & a_{24}e^{i\phi_{24}} \cr
\end{pmatrix}
\end{equation}
Eqs.~(\ref{eq:outputgate0}) and (\ref{eq:outputgate1})
become
\begin{eqnarray}
B[\phi_0]&=&\left(e^{i\varphi_1}\cos\vartheta B[0]
+e^{i\varphi_2}\sin\vartheta B[1]\right):= yU[0];
\label{eq:MPSappcond1}\\
B[\phi_1]&=&\left(e^{-i\varphi_2}\sin\vartheta B[0]
-e^{-i\varphi_1}\cos\vartheta B[1]\right):= yU[1],
\end{eqnarray}
where 
\begin{equation}
U[0]=\begin{pmatrix}
e^{i\phi_1}\cos\theta & e^{i\phi_2}\sin\theta\cr
e^{-i\phi_2}\sin\theta & -e^{-i\phi_1}\cos\theta\cr
\end{pmatrix}
\end{equation}
is an arbitrary target single-qubit unitary gate, and $U[1]$ and the constant
$y$ are to be determined. Solving Eq.~(\ref{eq:MPSappcond1}) yields constraints
on the parameters appearing in $B[1]$:
\begin{eqnarray}
a_{21}&=&e^{-i(\phi_{21}+\varphi_2)}\left(-a_{11}e^{i(\phi_{11}+\varphi_1)}
\cot\vartheta+y e^{i\phi_1}\cos\theta\csc\vartheta\right);\nonumber \\
a_{22}&=&e^{-i(\phi_{22}+\varphi_2)}\left(-a_{12}e^{i(\phi_{12}+\varphi_1)}
\cot\vartheta+y e^{i\phi_2}\sin\theta\csc\vartheta\right);\nonumber \\
a_{23}&=&e^{-i(\phi_{23}+\phi_2+\varphi_2)}\left(-a_{13}
e^{i(\phi_{13}+\varphi_1+\phi_2)}\cot\vartheta+y\sin\theta\csc\vartheta\right);
\nonumber \\
a_{24}&=&e^{-i(\phi_{24}+\phi_1+\varphi_2)}\left(-a_{14}
e^{i(\phi_{14}+\varphi_1+\phi_1)}\cot\vartheta-y\cos\theta\csc\vartheta\right).
\end{eqnarray}
The MPS matrices can be expressed in a column-oriented form similar to those of
the cluster state, Eq.~(\ref{eq:MPS4cluster}, by setting $a_{21}=a_{23}=0$,
which can be accomplished via
\begin{eqnarray}
a_{11}&=&y\cos\theta\sec\vartheta;\;\phi_{11}=\phi_1-\varphi_1;\;
a_{13}=y\sin\theta\sec\vartheta;\;\phi_{13}=-\phi_2-\varphi_1;\nonumber \\
a_{12}&=&0;\;\phi_{22}=\phi_2-\varphi_2;\;
a_{14}=0;\;\phi_{24}=-\phi_1-\varphi_2;
\end{eqnarray}
this yields
\begin{equation}
B[0]=ye^{-i\varphi_1}\sec\vartheta\begin{pmatrix}
e^{i\phi_1}\cos\theta & 0\cr
e^{-i\phi_2}\sin\theta & 0\cr
\end{pmatrix};\quad
B[1]=ye^{-i\varphi_2}\csc\vartheta\begin{pmatrix}
0 & e^{i\phi_2}\sin\theta\cr
0 & -e^{-i\phi_1}\cos\theta\cr
\end{pmatrix}.
\end{equation}
If the MPS matrices depend explicitly on all the measurement angles then MBQC
is impossible. Setting $\phi_1=\varphi_1$, $\phi_2=-\varphi_1$, and 
$\varphi_2=-\varphi_1$ yields
\begin{equation}
B[0]=y\sec\vartheta\begin{pmatrix}
\cos\theta & 0\cr
\sin\theta & 0\cr
\end{pmatrix};\quad
B[1]=y\csc\vartheta\begin{pmatrix}
0 & \sin\theta\cr
0 & -\cos\theta\cr
\end{pmatrix},
\end{equation}
with only $\vartheta$ remaining. The normalization condition (neglecting the 
junk sector) is
\begin{equation}
B[0]^{\dag}B[0]+B[1]^{\dag}B[1]
=y^2\begin{pmatrix}
\sec^2\vartheta & 0\cr
0 & \csc^2\vartheta\cr
\end{pmatrix}=I,
\end{equation}
which yields $\vartheta=(2k+1)\pi/4$, $k\in\mathbb{Z}$, and $y=1/\sqrt{2}$. 
Thus, the measurement angle $\vartheta$ is fixed; choosing $\vartheta=\pi/4$ 
one obtains
\begin{equation}
B[0]=\begin{pmatrix}
\cos\theta & 0\cr
\sin\theta & 0\cr
\end{pmatrix};\quad
B[1]=\begin{pmatrix}
0 & \sin\theta\cr
0 & -\cos\theta\cr
\end{pmatrix}.
\label{eq:MPS4new}
\end{equation}
The output unitary matrices are then
\begin{eqnarray}
U[0]&=&\begin{pmatrix}
e^{i\varphi_1}\cos\theta & e^{-i\varphi_1}\sin\theta\cr
e^{i\varphi_1}\sin\theta & -e^{-i\varphi_1}\cos\theta\cr
\end{pmatrix}=ZR_Y(\theta)R_Z(\varphi);\nonumber \\
U[1]&=&\begin{pmatrix}
e^{i\varphi_1}\cos\theta & -e^{-i\varphi_1}\sin\theta\cr
e^{i\varphi_1}\sin\theta & e^{-i\varphi_1}\cos\theta\cr
\end{pmatrix}=R_Y(-\theta)R_Z(\varphi)=ZR_Y(\theta)ZR_Z(\varphi)=U[0]Z,
\label{eq:U4new}
\end{eqnarray}
where $R_{\alpha}(\theta):=\exp(i\theta\alpha)$.

\section{SPT order}
\label{app:SPTorder}

In this Appendix we show that the states defined by Eq.~(\ref{eq:tensorstate1})
have non-trivial symmetry-protected topological order, by explicitly 
constructing the symmetry operators in real space.

\subsection{Review of SPT order in cluster states}

It is useful to review the basics of SPT order in cluster states, and the 
following analysis follows expands on Ref.~\onlinecite{Yoshida2016}. The 
$\mathbb{Z}_2\times\mathbb{Z}_2$ symmetry of the cluster state with an even
number of sites $n$ and periodic boundary conditions is explicitly generated by 
the operators $X_{\rm even}=\prod_jX_{2j}$ and $X_{\rm odd}=\prod_jX_{2j+1}$. 
Because $(X\otimes I)C_Z=C_Z(X\otimes Z)$, applying the $X_j$ operator on 
$C_Z^{j-1,j}C_Z^{j,j+1}$ returns $C_Z^{j-1,j}C_Z^{j,j+1}Z_{j-1}X_jZ_{j+1}$ so 
that all $Z$ factors cancel on the application of either $X_{\rm odd}$ or 
$X_{\rm even}$. The symmetry operators leave the cluster state invariant 
because all qubits are originally set to $|+\rangle$ which is an eigenstate of
$X$.

For a cluster state with open boundary conditions, the symmetry operators need
to be slightly modified. Again assume $n$ is even. The additional $Z$ operators 
resulting from the action of $X$ on the first and last sites, 
$X_1C_Z^{1,2}=C_Z^{1,2}X_1Z_2$ and $X_nC_Z^{n-1,n}=C_Z^{n-1,n}Z_{n-1}X_n$, will 
be cancelled by other $Z$ gates arising from the adjacent odd or even sites, 
respectively. But the action of $X$ on site 2 and $n-1$,
$X_2C_Z^{1,2}=C_Z^{1,2}Z_1X_2$ and $X_{n-1}C_Z^{n-1,n}=C_Z^{n-1,n}X_{n-1}Z_n$, 
yield $Z$ operators on the first and last sites that aren't cancelled by other
$X$ gates in $X_{\rm even}$ or $X_{\rm odd}$. The symmetry operators therefore
become $X_{\rm even}=Z_1\prod_jX_{2j}$ and $X_{\rm odd}=\prod_jX_{2j+1}Z_n$. 

The $C_Z$ gates are diagonal and therefore commute with $Z$ operators, so the 
cluster state is the unique $+n$ eigenstate of the $n$-fold sum
of stabilizer generators in the bulk $S_j=Z_{j-1}X_jZ_{j+1}$ 
($2\leq j\leq n-1$) and at the boundaries $S_1=X_1Z_2$ and $S_n=Z_{n-1}X_n$. 
Pauli gates at the boundaries are transformed by the $C_Z$ operators into 
effective Pauli gates $\overline{X}_1=C_Z^{1,2}X_1C_Z^{1,2}=X_1Z_2$ (note that 
$C_Z^{\dag}=C_Z$), 
$\overline{Z}_1=Z_1$, $\overline{X}_n=Z_{n-1}X_n$, and $\overline{Z}_n=Z_n$. 
Defining the generators of the $\mathbb{Z}_2\times\mathbb{Z}_2$ symmetry as 
$U(g_1,g_2)=X_{\rm odd}^{g_1}X_{\rm even}^{g_2}$, where $g_1,g_2\in\{0,1\}$,
the effective Pauli operators on the left site are transformed as 
$U(g_1,g_2)\overline{X}_1U(g_1,g_2)^{\dag}=(-1)^{g_2}\overline{X}_1$ and
$U(g_1,g_2)\overline{Z}_1U(g_1,g_2)^{\dag}=(-1)^{g_1}\overline{Z}_1$, which is
equivalent to an effective transformation $U_{\rm eff}(g_1,g_2)
=\overline{X}^{g_1}\overline{Z}^{g_2}$. A similar result holds for the right 
edge. 

SPT phases in one-dimensional systems can be classified by the second 
cohomology group. If $U_{\rm eff}$ is a projective representation of the 
symmetry group, then $U_{\rm eff}(g)U_{\rm eff}(h)
=\omega(g,h)U_{\rm eff}(g\oplus h)$, where
$\oplus$ here corresponds to bitwise addition (not a direct sum!) and
$\omega(g,h)$ (called a 2-cocycle) must satisfy the consistency 
conditions~\cite{Else2014}
\begin{equation}
\omega(g,h)\omega(g\oplus h,k)=\omega(h,k)
\omega(g,h\oplus k)
\label{eq:2cocycle}
\end{equation}
and
\begin{equation}
\omega(g,h)\sim\omega(g,h)\beta(g)\beta(h)
\beta(g\oplus h)^{-1},
\label{eq:2cocycle2}
\end{equation}
where the $\beta$ are phase factors. The 2-cocycle for the left boundary of the
cluster state is therefore
\begin{eqnarray}
\omega(g_1,g_2;h_1,h_2)&=&U_{\rm eff}(g_1,g_2)U_{\rm eff}(h_1,h_2)
U^{-1}_{\rm eff}(g_1\oplus h_1,g_2\oplus h_2)
=\overline{X}^{g_1}\overline{Z}^{g_2}
\overline{X}^{h_1}\overline{Z}^{h_2}\overline{Z}^{g_2\oplus h_2}
\overline{X}^{g_1\oplus h_1}\nonumber \\
&=&(-1)^{g_2h_1}I,
\label{eq:cocyclecluster}
\end{eqnarray}
where the result is obtained by considering each case. Check that this
satisfies Eq.~(\ref{eq:2cocycle}): $(-1)^{g_2h_1}(-1)^{(g_2\oplus h_2)k_1}
=(-1)^{h_2k_1}(-1)^{g_2(h_1\oplus k_1)}$. If $g_2=0$ then the left and right 
sides are $(-1)^{h_2k_1}$, and if $g_2=1$ then one requires 
$(-1)^{h_1}(-1)^{\bar{h}_2k_1}=(-1)^{h_2k_1}(-1)^{h_1\oplus k_1}$. Next, if 
$k_1=0$ then both sides are $(-1)^{h_1}$, and if $k_1=1$ then one requires 
$(-1)^{h_1}(-1)^{\bar{h}_2}=(-1)^{h_2}(-1)^{\bar{h}_1}$ or 
$(-1)^{h_1-\bar{h}_1}=(-1)^{h_2-\bar{h}_2}$ which is true for any 
$\{h_1,h_2\}$. Therefore Eq.~(\ref{eq:2cocycle}) is satisfied. A sufficient 
condition for SPT order is that the 2-cocycles anticommute:
\begin{eqnarray}
\omega(g_1,g_2;h_1,h_2)\omega(h_1,h_2;g_1,g_2)^{-1}&=&
U_{\rm eff}(g_1,g_2)U_{\rm eff}(h_1,h_2)U_{\rm eff}(g_1,g_2)^{-1}
U_{\rm eff}(h_1,h_2)^{-1}\nonumber \\
&=&(X_1Z_2)^{g_1}Z_1^{g_2}(X_1Z_2)^{h_1}Z_1^{h_2}Z_1^{g_2}(Z_2X_1)^{g_1}
Z_1^{h_2}(Z_2X_1)^{h_1}\nonumber \\
&=&X_1^{g_1}Z_1^{g_2}X_1^{h_1}Z_1^{h_2}Z_1^{g_2}X_1^{g_1}Z_1^{h_2}X_1^{h_1}
\nonumber \\
&=&X_1^{g_1}(-1)^{g_2h_1}X_1^{h_1}Z_1^{h_2}X_1^{g_1}Z_1^{h_2}X_1^{h_1}
\nonumber \\
&=&(-1)^{g_1h_2}(-1)^{g_2h_1}I\neq I,
\end{eqnarray}
consistent with Eq.~(\ref{eq:cocyclecluster}); thus, the algebra associated 
with the two effective Pauli operators at the surface is (maximally)
non-commutative~\cite{Miller2016}.

\subsection{Symmetry operators for the state defined in 
Eq.~(\ref{eq:tensorstate1})}

We first show that there exist two-qubit operators $P_{\theta_j}$ and 
$Q_{\theta_j}$, defined as
\begin{eqnarray}
P_{\theta_j}^{j,j+1}&=&|0\rangle_j\langle 0|X_{j+1}M_{\theta_j}^{j+1}X_{j+1}
+|1\rangle_j\langle 1|M_{\theta_j}^{j+1};\nonumber \\
Q_{\theta_j}^{j,j+1}&=&|0\rangle_j\langle 1|X_{j+1}M_{\theta_j}^{j+1}X_{j+1}
+|1\rangle_j\langle 0|M_{\theta_j}^{j+1}=P_{\theta_j}^{j,j+1}X_j,
\end{eqnarray}
where
\begin{equation}
M_{\theta_j}^{j+1}=\tan\theta_j|0\rangle_{j+1}\langle 0|
+\cot\theta_j|1\rangle_{j+1}\langle 1|,
\end{equation}
such that
\begin{equation}
P_{\theta_j}^{j-1,j}Q_{\theta_{j+1}}^{j,j+1}C_{\theta_j}^{j-1,j}
C_{\theta_{j+1}}^{j,j+1}
=P_{\theta_j}^{j-1,j}P_{\theta_{j+1}}^{j,j+1}X_jC_{\theta_j}^{j-1,j}
C_{\theta_{j+1}}^{j,j+1}
=C_{\theta_j}^{j-1,j}C_{\theta_{j+1}}^{j,j+1}
\left(Z_{j-1}X_jZ_{j+1}\right),
\label{eq:needtoshow}
\end{equation}
i.e.\ that
\begin{equation}
P_{\theta_j}^{j-1,j}Q_{\theta_{j+1}}^{j,j+1}=C_{\theta_j}^{j-1,j}
C_{\theta_{j+1}}^{j,j+1}\left(Z_{j-1}X_jZ_{j+1}\right)\left(
C_{\theta_j}^{j-1,j}C_{\theta_{j+1}}^{j,j+1}\right)^{-1}.
\end{equation}
Note that the inverse is applied since $C_{\theta}$ is not unitary.

Neglecting the factor of $\sqrt{2}$, the two-qubit operator 
$C_{\theta_j}^{j,j+1}$ defined in Eq.~(\ref{eq:Ctheta}) can be conveniently 
expressed as 
\begin{equation}
C_{\theta_j}^{j,j+1}=|0\rangle_j\langle 0|N_{\theta_j}^{j+1}
+|1\rangle_j\langle 1|X_{j+1}N_{\theta_j}^{j+1}X_{j+1}Z_{j+1},
\end{equation}
where
\begin{equation}
N_{\theta_j}^{j+1}=\cos\theta_j|0\rangle_{j+1}\langle 0|
+\sin\theta_j|1\rangle_{j+1}\langle 1|.
\end{equation}
To simplify the notation without loss of generality, rewrite
\begin{equation}
C_{\theta_j}^{j,j+1}\to C_j=|0\rangle\langle 0|\otimes N_j
+|1\rangle\langle 1|\otimes(XN_jXZ),
\end{equation}
with
\begin{equation}
N_j=\cos\theta_j|0\rangle\langle 0|+\sin\theta_j|1\rangle\langle 1|,
\end{equation}
and furthermore assume that $j=1$. A few lines of algebra yields
\begin{equation}
C_{\theta_j}^{j,j+1}C_{\theta_{j+1}}^{j+1,j+2}\to(C_1\otimes I)(I\otimes C_2)
=N_1\otimes|0\rangle\langle 0|\otimes N_2+XN_1XZ\otimes|1\rangle\langle 1|
\otimes XN_2XZ,
\end{equation}
and
\begin{equation}
(C_1\otimes I)(I\otimes C_2)(Z\otimes X\otimes Z)=N_1Z\otimes
|0\rangle\langle 1|\otimes N_2Z+XN_1X\otimes|1\rangle\langle 0|\otimes XN_2X.
\label{eq:rhs}
\end{equation}
Next, one seeks operators $P_{\theta_j}^{j,j+1}\to P_1\otimes I$ and 
$Q_{\theta_{j+1}}^{j+1,j+2}\to I\otimes Q_2$, where
\begin{eqnarray}
P_1&=&|0\rangle\langle 0|\otimes\alpha_1
+|0\rangle\langle 1|\otimes\alpha_2
+|1\rangle\langle 0|\otimes\alpha_3
+|1\rangle\langle 1|\otimes\alpha_4;\nonumber \\
Q_2&=&|0\rangle\langle 0|\otimes\beta_1
+|0\rangle\langle 1|\otimes\beta_2
+|1\rangle\langle 0|\otimes\beta_3
+|1\rangle\langle 1|\otimes\beta_4,
\end{eqnarray}
and the $\alpha_i$ and $\beta_i$ are free parameters,
such that $(P_1\otimes I)(I\otimes Q_2)(C_1\otimes I)(I\otimes C_2)$ 
returns the right hand side of Eq.~(\ref{eq:rhs}). Straightforward 
algebra yields
\begin{eqnarray}
(P_1\otimes I)(I\otimes Q_2)(C_1\otimes I)(I\otimes C_2)&=&
\cos\theta_1|0\rangle\langle 0|\otimes\alpha_1|0\rangle\langle 0|\otimes\beta_1
N_2+\cos\theta_1|0\rangle\langle 0|\otimes\alpha_1|1\rangle\langle 0|\otimes
\beta_3N_2\nonumber \\
&+&\sin\theta_1|0\rangle\langle 0|\otimes\alpha_1|0\rangle\langle 1|\otimes
\beta_2XN_2XZ+\sin\theta_1|0\rangle\langle 0|\otimes\alpha_1|1\rangle\langle 1|
\otimes\beta_4XN_2XZ\nonumber \\
&+&\sin\theta_1|1\rangle\langle 1|\otimes\alpha_4|0\rangle\langle 0|\otimes
\beta_1N_2+\sin\theta_1|1\rangle\langle 1|\otimes\alpha_4|1\rangle\langle 0|
\otimes\beta_3N_2\nonumber \\
&-&\cos\theta_1|1\rangle\langle 1|\otimes\alpha_4|0\rangle\langle 1|\otimes
\beta_2XN_2XZ-\cos\theta_1|1\rangle\langle 1|\otimes\alpha_4|1\rangle\langle 1|
\otimes\beta_4XN_2XZ,\nonumber \\
\end{eqnarray}
and $\alpha_2=\alpha_3=0$. Comparing this expression with the right hand side
of Eq.~(\ref{eq:rhs}), the free parameters must take the values
\begin{eqnarray}
\alpha_4&=&\tan\theta_1|0\rangle\langle 0|+\cot\theta_1|1\rangle\langle 1|;
\quad\alpha_1=\cot\theta_1|0\rangle\langle 0|+\tan\theta_1|1\rangle\langle 1|
=X\alpha_4X;\nonumber \\
\beta_3&=&\tan\theta_2|0\rangle\langle 0|+\cot\theta_2|1\rangle\langle 1|;
\quad\beta_2=\cot\theta_2|0\rangle\langle 0|+\tan\theta_2|1\rangle\langle 1|
=X\beta_3X;\quad\beta_1=\beta_4=0.
\end{eqnarray}
One then obtains
\begin{eqnarray}
P_1&=&|0\rangle\langle 0|\otimes XM_1X+|1\rangle\langle 1|\otimes M_1;
\nonumber \\
Q_2&=&|0\rangle\langle 1|\otimes XM_2X+|1\rangle\langle 0|\otimes M_2,
\label{eq:AB}
\end{eqnarray}
where
\begin{equation}
M_j=\tan\theta_j|0\rangle\langle 0|+\cot\theta_j|1\rangle\langle 1|.
\label{eq:M}
\end{equation}
Reverting to unsimplified notation, one obtains
\begin{equation}
P_{\theta_j}^{j,j+1}Q_{\theta_{j+1}}^{j+1,j+2}C_{\theta_j}^{j,j+1}
C_{\theta_{j+1}}^{j+1,j+2}
=P_{\theta_j}^{j,j+1}P_{\theta_{j+1}}^{j+1,j+2}X_{j+1}C_{\theta_j}^{j,j+1}
C_{\theta_{j+1}}^{j+1,j+2}
=C_{\theta_j}^{j,j+1}C_{\theta_{j+1}}^{j+1,j+2}\left(Z_jX_{j+1}Z_{j+2}\right),
\label{eq:genstab1}
\end{equation}
where
\begin{equation}
P_{\theta_j}^{j,j+1}=|0\rangle_j\langle 0|X_{j+1}M_{\theta_j}^{j+1}X_{j+1}
+|1\rangle_j\langle 1|M_{\theta_j}^{j+1}
\label{eq:Atheta}
\end{equation}
and
\begin{equation}
M_{\theta_j}^{j+1}=\tan\theta_j|0\rangle_{j+1}\langle 0|
+\cot\theta_j|1\rangle_{j+1}\langle 1|,
\end{equation}
though the non-unitary operator $P_{\theta_j}^{j,j+1}$ is more conveniently 
expressed as a diagonal matrix acting on qubits $j$ and $j+1$:
\begin{equation}
P_{\theta_j}^{j,j+1}=\mbox{diag}(\cot\theta_j,\tan\theta_j,\tan\theta_j,
\cot\theta_j)_{j,j+1}.
\label{eq:Pdiag}
\end{equation}
These recover the statement Eq.~(\ref{eq:needtoshow}). 
Using the same procedure, one may derive similar expressions for the boundary
operators:
\begin{equation}
P_{\theta_1}^{1,2}X_1C_{\theta_1}^{1,2}C_{\theta_2}^{2,3}
=C_{\theta_1}^{1,2}C_{\theta_2}^{2,3}\left(X_1Z_2I_3\right);\quad
P_{\theta_n}^{n-1,n}X_nC_{\theta_{n-1}}^{n-2,n-1}C_{\theta_n}^{n-1,n}
=C_{\theta_{n-1}}^{n-2,n-1}C_{\theta_n}^{n-1,n}
\left(I_{n-2}Z_{n-1}X_n\right).
\label{eq:boundaryops}
\end{equation}

Second, we prove that the joint operator $P_{\theta_j}^{j,j+1}
Q_{\theta_{j+1}}^{j+1,j+2}$ commutes with its counterpart 
$P_{\theta_{j+2}}^{j+2,j+3}Q_{\theta_{j+3}}^{j+3,j+4}$ two sites over. Using
the simplified notation, this corresponds to proving that $[(P_1\otimes I)
(I\otimes Q_2)\otimes I^{\otimes 2},I^{\otimes 2}\otimes(P_3\otimes I)
(I\otimes Q_4)]=0$. The third qubit is the only one with support on both 
operators, so one need only focus on the contributions of the operators on this
position: the $XM_2X$ and $M_2$ from $Q_2$, and the $|0\rangle\langle 0|$ and 
$|1\rangle\langle 1|$ from $P_3$. And, because all of these operators are 
diagonal in the computational basis, they commute, and therefore the 
$P_{\theta_j}^{j,j+1}Q_{\theta_{j+1}}^{j+1,j+2}$ and
$P_{\theta_{j+2}}^{j+2,j+3}Q_{\theta_{j+3}}^{j+3,j+4}$ operators commute. Thus, 
$\prod_jP_{\theta_{2j}}^{2j,2j+1}Q_{\theta_{2j+1}}^{2j+1,2j+2}
=\prod_jP_{\theta_{2j}}^{2j,2j+1}P_{\theta_{2j+1}}^{2j+1,2j+2}X_{2j+1}$ 
and 
$\prod_jP_{\theta_{2j-1}}^{2j-1,2j}Q_{\theta_{2j}}^{2j,2j+1}
=\prod_jP_{\theta_{2j-1}}^{2j-1,2j}P_{\theta_{2j}}^{2j,2j+1}X_{2j}$ are
bulk symmetry operators for the state (\ref{eq:tensorstate1}). Likewise, the 
boundary operators in 
Eq.~(\ref{eq:boundaryops}) automatically commute with the bulk operators two 
sites over, because they have support on different qubits. 
The analogs of the cluster-state symmetry operators (for even $n$) are then
\begin{equation}
X_{\rm odd}=P_{\theta_1}^{1,2}X_1\left(\prod_{j=1}
P_{\theta_{2j}}^{2j,2j+1}P_{\theta_{2j+1}}^{2j+1,2j+2}X_{2j+1}\right)Z_n;
\quad X_{\rm even}=Z_1\left(\prod_{j=1}P_{\theta_{2j-1}}^{2j-1,2j}
P_{\theta_{2j}}^{2j,2j+1}X_{2j}\right)\left(P_{\theta_n}^{n-1,n}X_n\right),
\label{eq:symmetryops}
\end{equation}
where $Z$ gates are added to the last and first sites of $X_{\rm odd}$ and 
$X_{\rm even}$, respectively, as was necessary for the cluster state with open
boundary conditions. Note that $X_{\rm odd}$ and $X_{\rm even}$ are neither 
unitary nor onsite global symmetries.

\subsection{Generalized stabilizers for the state defined in 
Eq.~(\ref{eq:tensorstate1})}

We first show that the $P_{\theta_j}$ operators yield $n-2$ operators 
$S_{j,j+1}$, $2\leq j\leq n-1$, such that $S_{j,j+1}|\psi\rangle=|\psi\rangle$. 
These are the analogs of the bulk cluster-state stabilizer generators 
$Z_{j-1}X_jZ_{j+1}$. Given that the $C_{\theta_j}$ gates are diagonal, one may 
rewrite Eq.~(\ref{eq:needtoshow}):
\begin{equation}
P_{\theta_j}^{j-1,j}P_{\theta_{j+1}}^{j,j+1}Z_{j-1}X_jZ_{j+1}
C_{\theta_j}^{j-1,j}C_{\theta_{j+1}}^{j,j+1}=S_{j,j+1}C_{\theta_j}^{j-1,j}
C_{\theta_{j+1}}^{j,j+1}=C_{\theta_j}^{j-1,j}C_{\theta_{j+1}}^{j,j+1}X_j.
\label{eq:genstabs1}
\end{equation}
Thus, $S_{j,j+1}=P_{\theta_j}^{j-1,j}P_{\theta_{j+1}}^{j,j+1}Z_{j-1}X_jZ_{j+1}$
are eigenoperators for the state with unit eigenvalue, for any $j$ in the bulk,
and generalize the stabilizer operators for the cluster state. With 
Eq.~(\ref{eq:Atheta}), one obtains:
\begin{eqnarray}
S_{j,j+1}&=&\left[|0\rangle_{j-1}\langle 0|X_jM_{\theta_j}^jX_j
+|1\rangle_{j-1}\langle 1|M_{\theta_j}^j\right]\left[|0\rangle_j\langle 0|
X_{j+1}M_{\theta_{j+1}}^{j+1}X_{j+1}+|1\rangle_j\langle 1|
M_{\theta_{j+1}}^{j+1}\right]Z_{j-1}X_jZ_{j+1}\nonumber \\
&=&\left[|0\rangle_{j-1}\langle 0|X_j\left(\tan\theta_j|0\rangle_j\langle 0|
+\cot\theta_j|1\rangle_j\langle 1|\right)X_j
+|1\rangle_{j-1}\langle 1|\left(\tan\theta_j|0\rangle_j\langle 0|
+\cot\theta_j|1\rangle_j\langle 1|\right)\right]\nonumber \\
&\times&\left[|0\rangle_j\langle 0|X_{j+1}\left(\tan\theta_{j+1}
|0\rangle_{j+1}\langle 0|+\cot\theta_{j+1}|1\rangle_{j+1}\langle 1|\right)
X_{j+1}+|1\rangle_j\langle 1|\left(\tan\theta_{j+1}|0\rangle_{j+1}\langle 0|
+\cot\theta_{j+1}|1\rangle_{j+1}\langle 1|\right)\right]\nonumber \\
&\times&Z_{j-1}X_jZ_{j+1}\nonumber \\
&=&\left[|0\rangle_{j-1}\langle 0|\left(\tan\theta_j|1\rangle_j\langle 1|
+\cot\theta_j|0\rangle_j\langle 0|\right)
+|1\rangle_{j-1}\langle 1|\left(\tan\theta_j|0\rangle_j\langle 0|
+\cot\theta_j|1\rangle_j\langle 1|\right)\right]\nonumber \\
&\times&\left[|0\rangle_j\langle 0|\left(\tan\theta_{j+1}
|1\rangle_{j+1}\langle 1|+\cot\theta_{j+1}|0\rangle_{j+1}\langle 0|\right)
+|1\rangle_j\langle 1|\left(\tan\theta_{j+1}|0\rangle_{j+1}\langle 0|
+\cot\theta_{j+1}|1\rangle_{j+1}\langle 1|\right)\right]Z_{j-1}X_jZ_{j+1}
\nonumber \\
&=&\left[|0\rangle_{j-1}\langle 0|\left(\cot\theta_j|0\rangle_j\langle 0|\right)
+|1\rangle_{j-1}\langle 1|\left(\tan\theta_j|0\rangle_j\langle 0|
\right)\right]\nonumber \\
&\times&\left[|0\rangle_j\langle 0|\left(\tan\theta_{j+1}
|1\rangle_{j+1}\langle 1|+\cot\theta_{j+1}|0\rangle_{j+1}\langle 0|\right)
\right]Z_{j-1}X_jZ_{j+1}
\nonumber \\
&+&\left[|0\rangle_{j-1}\langle 0|\left(\tan\theta_j|1\rangle_j\langle 1|
\right)
+|1\rangle_{j-1}\langle 1|\left(\cot\theta_j|1\rangle_j\langle 1|\right)
\right]\nonumber \\
&\times&\left[|1\rangle_j\langle 1|\left(\tan\theta_{j+1}|0\rangle_{j+1}
\langle 0|+\cot\theta_{j+1}|1\rangle_{j+1}\langle 1|\right)\right]Z_{j-1}
X_jZ_{j+1}
\nonumber \\
&=&\left(\cot\theta_j|0\rangle_{j-1}\langle 0|+\tan\theta_j|1\rangle_{j-1}
\langle 1|\right)|0\rangle_j\langle 0|\left(\tan\theta_{j+1}
|1\rangle_{j+1}\langle 1|+\cot\theta_{j+1}|0\rangle_{j+1}\langle 0|\right)
Z_{j-1}X_jZ_{j+1}\nonumber \\
&+&\left(\tan\theta_j|0\rangle_{j-1}\langle 0|
+\cot\theta_j|1\rangle_{j-1}\langle 1|\right)|1\rangle_j\langle 1|
\left(\tan\theta_{j+1}|0\rangle_{j+1}\langle 0|+\cot\theta_{j+1}|1\rangle_{j+1}
\langle 1|\right)Z_{j-1}X_jZ_{j+1}\nonumber \\
&=&\left(\cot\theta_j|0\rangle_{j-1}\langle 0|-\tan\theta_j|1\rangle_{j-1}
\langle 1|\right)|0\rangle_j\langle 1|\left(-\tan\theta_{j+1}
|1\rangle_{j+1}\langle 1|+\cot\theta_{j+1}|0\rangle_{j+1}\langle 0|\right)
\nonumber \\
&+&\left(\tan\theta_j|0\rangle_{j-1}\langle 0|
-\cot\theta_j|1\rangle_{j-1}\langle 1|\right)|1\rangle_j\langle 0|
\left(\tan\theta_{j+1}|0\rangle_{j+1}\langle 0|-\cot\theta_{j+1}|1\rangle_{j+1}
\langle 1|\right).
\label{eq:genstabs1}
\end{eqnarray}
Note that the $S_{j,j+1}$ are non-separable three-local operators that are
neither unitary nor Hermitian. These operators square to the identity, as 
required for generalied stabilizers:
\begin{eqnarray}
S_{j,j+1}^2&=&\big[\left(\cot\theta_j|0\rangle_{j-1}\langle 0|-\tan\theta_j|1\rangle_{j-1}
\langle 1|\right)|0\rangle_j\langle 1|\left(-\tan\theta_{j+1}
|1\rangle_{j+1}\langle 1|+\cot\theta_{j+1}|0\rangle_{j+1}\langle 0|\right)
\nonumber \\
&+&\left(\tan\theta_j|0\rangle_{j-1}\langle 0|
-\cot\theta_j|1\rangle_{j-1}\langle 1|\right)|1\rangle_j\langle 0|
\left(\tan\theta_{j+1}|0\rangle_{j+1}\langle 0|-\cot\theta_{j+1}|1\rangle_{j+1}
\langle 1|\right)\big]\nonumber \\
&\times&\big[\left(\cot\theta_j|0\rangle_{j-1}\langle 0|-\tan\theta_j
|1\rangle_{j-1}\langle 1|\right)|0\rangle_j\langle 1|\left(-\tan\theta_{j+1}
|1\rangle_{j+1}\langle 1|+\cot\theta_{j+1}|0\rangle_{j+1}\langle 0|\right)
\nonumber \\
&+&\left(\tan\theta_j|0\rangle_{j-1}\langle 0|
-\cot\theta_j|1\rangle_{j-1}\langle 1|\right)|1\rangle_j\langle 0|
\left(\tan\theta_{j+1}|0\rangle_{j+1}\langle 0|-\cot\theta_{j+1}|1\rangle_{j+1}
\langle 1|\right)\big]\nonumber \\
&=&\big[\left(\cot\theta_j|0\rangle_{j-1}\langle 0|-\tan\theta_j|1\rangle_{j-1}
\langle 1|\right)|0\rangle_j\langle 1|\left(-\tan\theta_{j+1}
|1\rangle_{j+1}\langle 1|+\cot\theta_{j+1}|0\rangle_{j+1}\langle 0|\right)
\big]\nonumber \\
&\times&\big[\left(\cot\theta_j|0\rangle_{j-1}\langle 0|-\tan\theta_j
|1\rangle_{j-1}\langle 1|\right)|0\rangle_j\langle 1|\left(-\tan\theta_{j+1}
|1\rangle_{j+1}\langle 1|+\cot\theta_{j+1}|0\rangle_{j+1}\langle 0|\right)
\big]\nonumber \\
&+&\big[\left(\cot\theta_j|0\rangle_{j-1}\langle 0|-\tan\theta_j|1\rangle_{j-1}
\langle 1|\right)|0\rangle_j\langle 1|\left(-\tan\theta_{j+1}
|1\rangle_{j+1}\langle 1|+\cot\theta_{j+1}|0\rangle_{j+1}\langle 0|\right)
\big]\nonumber \\
&\times&\big[\left(\tan\theta_j|0\rangle_{j-1}\langle 0|
-\cot\theta_j|1\rangle_{j-1}\langle 1|\right)|1\rangle_j\langle 0|
\left(\tan\theta_{j+1}|0\rangle_{j+1}\langle 0|-\cot\theta_{j+1}|1\rangle_{j+1}
\langle 1|\right)\big]\nonumber \\
&+&\big[\left(\tan\theta_j|0\rangle_{j-1}\langle 0|
-\cot\theta_j|1\rangle_{j-1}\langle 1|\right)|1\rangle_j\langle 0|
\left(\tan\theta_{j+1}|0\rangle_{j+1}\langle 0|-\cot\theta_{j+1}|1\rangle_{j+1}
\langle 1|\right)\big]\nonumber \\
&\times&\big[\left(\cot\theta_j|0\rangle_{j-1}\langle 0|-\tan\theta_j
|1\rangle_{j-1}\langle 1|\right)|0\rangle_j\langle 1|\left(-\tan\theta_{j+1}
|1\rangle_{j+1}\langle 1|+\cot\theta_{j+1}|0\rangle_{j+1}\langle 0|\right)
\big]\nonumber \\
&+&\big[\left(\tan\theta_j|0\rangle_{j-1}\langle 0|
-\cot\theta_j|1\rangle_{j-1}\langle 1|\right)|1\rangle_j\langle 0|
\left(\tan\theta_{j+1}|0\rangle_{j+1}\langle 0|-\cot\theta_{j+1}|1\rangle_{j+1}
\langle 1|\right)\big]\nonumber \\
&\times&\big[\left(\tan\theta_j|0\rangle_{j-1}\langle 0|
-\cot\theta_j|1\rangle_{j-1}\langle 1|\right)|1\rangle_j\langle 0|
\left(\tan\theta_{j+1}|0\rangle_{j+1}\langle 0|-\cot\theta_{j+1}|1\rangle_{j+1}
\langle 1|\right)\big]\nonumber \\
&=&\big[\left(\cot\theta_j|0\rangle_{j-1}\langle 0|-\tan\theta_j|1\rangle_{j-1}
\langle 1|\right)|0\rangle_j\langle 1|\left(-\tan\theta_{j+1}
|1\rangle_{j+1}\langle 1|+\cot\theta_{j+1}|0\rangle_{j+1}\langle 0|\right)
\big]\nonumber \\
&\times&\big[\left(\tan\theta_j|0\rangle_{j-1}\langle 0|
-\cot\theta_j|1\rangle_{j-1}\langle 1|\right)|1\rangle_j\langle 0|
\left(\tan\theta_{j+1}|0\rangle_{j+1}\langle 0|-\cot\theta_{j+1}|1\rangle_{j+1}
\langle 1|\right)\big]\nonumber \\
&+&\big[\left(\tan\theta_j|0\rangle_{j-1}\langle 0|
-\cot\theta_j|1\rangle_{j-1}\langle 1|\right)|1\rangle_j\langle 0|
\left(\tan\theta_{j+1}|0\rangle_{j+1}\langle 0|-\cot\theta_{j+1}|1\rangle_{j+1}
\langle 1|\right)\big]\nonumber \\
&\times&\big[\left(\cot\theta_j|0\rangle_{j-1}\langle 0|-\tan\theta_j
|1\rangle_{j-1}\langle 1|\right)|0\rangle_j\langle 1|\left(-\tan\theta_{j+1}
|1\rangle_{j+1}\langle 1|+\cot\theta_{j+1}|0\rangle_{j+1}\langle 0|\right)
\big]\nonumber \\
&=&\big[\left(|0\rangle_{j-1}\langle 0|+|1\rangle_{j-1}\langle 1|\right)
|0\rangle_j\langle 0|\left(|1\rangle_{j+1}\langle 1|+|0\rangle_{j+1}\langle 0|
\right)\big]\nonumber \\
&+&\big[\left(|0\rangle_{j-1}\langle 0|+|1\rangle_{j-1}\langle 1|\right)
|1\rangle_j\langle 1|\left(|0\rangle_{j+1}\langle 0|+|1\rangle_{j+1}\langle 1|
\right)\big]=I_{j-1,j,j+1},
\label{eq:Ssquare}
\end{eqnarray}
as desired. 

Second, internal consistency also requires that the generalized stabilizer 
operators for different $j$ values commute. In simplified notation, one needs 
to only verify that $[S_{1,2}\otimes I,I\otimes S_{2,3}]=0$, where using 
Eqs.~(\ref{eq:AB}) and (\ref{eq:M}) one obtains
\begin{eqnarray}
S_{j,j+1}&=&\cot\theta_j|0\rangle\langle 0|\otimes|0\rangle\langle 1|
\otimes XM_{j+1}XZ-\tan\theta_j|1\rangle\langle 1|\otimes|0\rangle\langle 1|
\otimes XM_{j+1}XZ\nonumber \\
&+&\tan\theta_j|0\rangle\langle 0|\otimes|1\rangle\langle 0|\otimes M_{j+1}Z
-\cot\theta_j|1\rangle\langle 1|\otimes|1\rangle\langle 0|\otimes M_{j+1}Z
\end{eqnarray}
where $j=1,2$ is strictly a label and does not denote qubit position.
Multiplication yields the unenlightening expressions
\begin{eqnarray}
(S_{1,2}\otimes I)(I\otimes S_{2,3})&=&-\cot\theta_1\tan\theta_2
|0\rangle\langle 0|\otimes|0\rangle\langle 1|\otimes XM_2XZ|0\rangle\langle 1|
\otimes XM_3XZ\nonumber \\
&-&\cot\theta_1\cot\theta_2|0\rangle\langle 0|\otimes|0\rangle\langle 1|
\otimes XM_2XZ|1\rangle\langle 0|\otimes M_3Z\nonumber \\
&+&\tan\theta_1\tan\theta_2|1\rangle\langle 1|\otimes|0\rangle\langle 1|
\otimes XM_2XZ|0\rangle\langle 1|\otimes XM_3XZ\nonumber \\
&+&\tan\theta_1\cot\theta_2|1\rangle\langle 1|\otimes|0\rangle\langle 1|
\otimes XM_2XZ|1\rangle\langle 0|\otimes M_3Z\nonumber \\
&+&\tan\theta_1\tan\theta_2|0\rangle\langle 0|\otimes|1\rangle\langle 0|
\otimes M_2Z|1\rangle\langle 0|\otimes M_3Z\nonumber \\
&+&\tan\theta_1\cot\theta_2|0\rangle\langle 0|\otimes|1\rangle\langle 0|
\otimes M_2Z|0\rangle\langle 1|\otimes XM_3XZ\nonumber \\
&-&\cot\theta_1\tan\theta_2|1\rangle\langle 1|\otimes|1\rangle\langle 0|
\otimes M_2Z|1\rangle\langle 0|\otimes M_3Z\nonumber \\
&-&\cot\theta_1\cot\theta_2|1\rangle\langle 1|\otimes|1\rangle\langle 0|
\otimes M_2Z|0\rangle\langle 1|\otimes XM_3XZ;\nonumber \\
(I\otimes S_{2,3})(S_{1,2}\otimes I)&=&\cot\theta_1\cot\theta_2
|0\rangle\langle 0|\otimes|0\rangle\langle 1|\otimes|0\rangle\langle 1|XM_2XZ
\otimes XM_3XZ\nonumber \\
&+&\cot\theta_1\tan\theta_2|0\rangle\langle 0|\otimes|0\rangle\langle 1|
\otimes|1\rangle\langle 0|XM_2XZ\otimes M_3Z\nonumber \\
&-&\tan\theta_1\cot\theta_2|1\rangle\langle 1|\otimes|0\rangle\langle 1|
\otimes|0\rangle\langle 1|XM_2XZ\otimes XM_3XZ\nonumber \\
&-&\tan\theta_1\tan\theta_2|1\rangle\langle 1|\otimes|0\rangle\langle 1|
\otimes|1\rangle\langle 0|XM_2XZ\otimes M_3Z\nonumber \\
&-&\tan\theta_1\cot\theta_2|0\rangle\langle 0|\otimes|1\rangle\langle 0|
\otimes|1\rangle\langle 0|M_2Z\otimes M_3Z\nonumber \\
&-&\tan\theta_1\tan\theta_2|0\rangle\langle 0|\otimes|1\rangle\langle 0|
\otimes|0\rangle\langle 1|M_2Z\otimes XM_3XZ\nonumber \\
&+&\cot\theta_1\cot\theta_2|1\rangle\langle 1|\otimes|1\rangle\langle 0|
\otimes|1\rangle\langle 0|M_2Z\otimes M_3Z\nonumber \\
&+&\cot\theta_1\tan\theta_2|1\rangle\langle 1|\otimes|1\rangle\langle 0|
\otimes|0\rangle\langle 1|M_2Z\otimes XM_3XZ.
\end{eqnarray}
As $XM_2XZ|0\rangle\langle 1|=\cot\theta_2|0\rangle\langle 1|$ while
$|0\rangle\langle 1|XM_2XZ=-\tan\theta_2|0\rangle\langle 1|$, terms 1 and 3
in both expressions coincide; similarly, as $XM_2XZ|1\rangle\langle 0|
=-\tan\theta_2|1\rangle\langle 0|$ while
$|1\rangle\langle 0|XM_2XZ=\cot\theta_2|1\rangle\langle 0|$, terms 2 and 4
in both expressions coincide. Similar results apply to the remaining terms, and 
therefore $[S_{1,2}\otimes I,I\otimes S_{2,3}]=0$. 

Third, one may likewise define generalizations of the surface stabilizers from 
the operations in Eq.~(\ref{eq:boundaryops}):
\begin{eqnarray}
P_{\theta_1}^{1,2}X_1Z_2C_{\theta_1}^{1,2}C_{\theta_2}^{2,3}
&=&S_{1,2}C_{\theta_1}^{1,2}C_{\theta_2}^{2,3}
=C_{\theta_1}^{1,2}C_{\theta_2}^{2,3}X_1;\nonumber \\
P_{\theta_n}^{n-1,n}Z_{n-1}X_nC_{\theta_{n-1}}^{n-2,n-1}C_{\theta_n}^{n-1,n}
&=&S_{n-1,n}C_{\theta_{n-1}}^{n-2,n-1}C_{\theta_n}^{n-1,n}
=C_{\theta_{n-1}}^{n-2,n-1}C_{\theta_n}^{n-1,n}X_n,
\label{eq:genstabs2}
\end{eqnarray}
so that $S_{1,2}=P_{\theta_1}^{1,2}X_1Z_2$ and 
$S_{n-1,n}=P_{\theta_n}^{n-1,n}Z_{n-1}X_n$. Following a similar analysis as 
above, it is straightforward to show that these operators also commute with the 
bulk generalized stabilizer generators, Eq.~(\ref{eq:genstabs1}). Likewise,
$S_{1,2}^2=I_{1,2}$ and $S_{n-1,n}^2=I_{n-1,n}$.

Fourth, for $X_{\rm odd}$ and $X_{\rm even}$, Eq.~(\ref{eq:symmetryops}), to 
represent a $\mathbb{Z}_2\times\mathbb{Z}_2$ symmetry, they should also square 
to the identity. We can make use of the results above for this purpose. Because 
$P_{\theta_j}^{j-1,j}$, Eq.~(\ref{eq:Pdiag}), is diagonal, one can write 
\begin{equation}
S_{j,j+1}=P_{\theta_j}^{j-1,j}P_{\theta_{j+1}}^{j,j+1}Z_{j-1}X_jZ_{j+1}
=Z_{j-1}Z_{j+1}P_{\theta_j}^{j-1,j}P_{\theta_{j+1}}^{j,j+1}X_j,
\end{equation} 
or
\begin{equation}
S_{2j+1,2j+2}=Z_{2j}Z_{2j+2}P_{\theta_{2j+1}}^{2j,2j+1}
P_{\theta_{2j+2}}^{2j+1,2j+2}X_{2j+1}=P_{\theta_{2j+1}}^{2j,2j+1}
P_{\theta_{2j+2}}^{2j+1,2j+2}X_{2j+1}Z_{2j}Z_{2j+2},
\end{equation}
so that 
\begin{equation}
P_{\theta_{2j+1}}^{2j,2j+1}P_{\theta_{2j+2}}^{2j+1,2j+2}X_{2j+1}=
Z_{2j}Z_{2j+2}S_{2j+1,2j+2}=S_{2j+1,2j+2}Z_{2j}Z_{2j+2}.
\label{eq:ZcommuteS}
\end{equation}
One can therefore rewrite $X_{\rm odd}$ as
\begin{eqnarray}
X_{\rm odd}&=&P_{\theta_1}^{1,2}X_1\left(\prod_{j=1}
P_{\theta_{2j}}^{2j,2j+1}P_{\theta_{2j+1}}^{2j+1,2j+2}X_{2j+1}\right)Z_n
=\left(Z_2S_{1,2}\right)\left(\prod_{j=1}Z_{2j}Z_{2j+2}S_{2j+1,2j+2}\right)
Z_n\nonumber \\
&=&\left(S_{1,2}\right)\left(\prod_{j=1}S_{2j+1,2j+2}\right),
\end{eqnarray}
where $\theta_{2j+1}\to\theta_{2j}$ is an unimportant shift; here we have used
the fact that all the $Z$ operators commute through the $S$ operators,
Eq.~(\ref{eq:ZcommuteS}).
Then
\begin{equation}
X_{\rm odd}^2=\left(S_{1,2}\right)\left(\prod_{j=1}S_{2j+1,2j+2}\right)
\left(S_{1,2}\right)\left(\prod_{j'=1}S_{2j'+1,2j'+2}\right)=I,
\end{equation}
because all generalized stabilizers commute with one another, as shown above, 
and then square to the identity, Eq.~(\ref{eq:Ssquare}). Thus, the symmetry 
operator $X_{\rm odd}$ squares to the identity. A similar result follows for 
$X_{\rm even}=\left(\prod_jS_{2j,2j+1}\right)\left(S_{n-1,n}\right)$. The 
symmetry operators written this way have an intuitive form, as products of 
generalized stabilizers on two-site blocks, counting either from the first or 
second site. It is important to keep in mind, however, that each generalized
stabilizer operator acts on three sites.

\subsection{SPT order for the state defined in Eq.~(\ref{eq:tensorstate1})}

Finally, one can apply these results to the analysis of SPT order. Again, we 
can treat the set of $n$ $S_{j,j+1}$ operators as effective stabilizers that 
uniquely define the state, including the qubits at the boundaries. The $X$ and 
$Z$ gates on the boundary qubit are transformed by the $C_{\theta}$ operators 
into effective Pauli gates, and can be read directly from 
Eq.~(\ref{eq:boundaryops}: 
\begin{equation}
\overline{X}_1=P_{\theta_1}^{1,2}X_1Z_2;\;\overline{Z}_1=Z_1;\;
\overline{X}_n=P_{\theta_n}^{n-1,n}Z_{n-1}X_n;\;\overline{Z}_n=Z_n.
\label{eq:effectivePaulis}
\end{equation}
Define the generators of the $\mathbb{Z}_2\times\mathbb{Z}_2$ as 
$O(g_1,g_2)=X_{\rm odd}^{g_1}X_{\rm even}^{g_2}$, where $X_{\rm even}$ and
$X_{\rm odd}$ are given in Eq.~(\ref{eq:symmetryops}) and $g_1,g_2\in\{0,1\}$.
We write $O(g_1,g_2)$ rather than $U(g_1,g_2)$ as the former is only unitary
for the specific case of $\theta_i=\theta_c$ $\forall i$.
The goal is to determine the effective Pauli operators $\overline{X}_1'$ and 
$\overline{Z}_1'$ for the left side that satisfy $\overline{X}_1'O(g_1,g_2)
=O(g_1,g_2)\overline{X}_1$ and $\overline{Z}_1'O(g_1,g_2)
=O(g_1,g_2)\overline{Z}_1$; recall that $O(g_1,g_2)^2=I$ so that 
$O(g_1,g_2)^{-1}=O(g_1,g_2)$. The latter is simpler:
\begin{equation}
\overline{Z}_1'O(g_1,g_2)=O(g_1,g_2)\overline{Z}_1
=\left(P_{\theta_1}^{1,2}X_1\right)^{g_1}\left(P_{\theta_1}^{1,2}
P_{\theta_2}^{2,3}X_2\right)^{g_2}Z_1
=Z_1(-1)^{g_1}\left(P_{\theta_1}^{1,2}X_1\right)^{g_1}
\left(P_{\theta_1}^{1,2}P_{\theta_2}^{2,3}X_2\right)^{g_2},
\end{equation}
which gives $\overline{Z}_1'=(-1)^{g_1}\overline{Z}_1$. Consider the cases
$O(1,0)\overline{X}_1$ and $O(0,1)\overline{X}_1$ separately:
\begin{eqnarray}
O(1,0)\overline{X}_1&=&P_{\theta_1}^{1,2}X_1P_{\theta_1}^{1,2}X_1Z_2
=(|0\rangle_1\langle 0|X_2M_{\theta_1}^2X_2
+|1\rangle_1\langle 1|M_{\theta_1}^2)(|1\rangle_1\langle 1|X_2M_{\theta_1}^2
X_2Z_2+|0\rangle_1\langle 0|M_{\theta_1}^2Z_2)\nonumber \\
&=&(|0\rangle_1\langle 0|X_2M_{\theta_1}^2X_2M_{\theta_1}^2Z_2
+|1\rangle_1\langle 1|M_{\theta_1}^2X_2M_{\theta_1}^2X_2Z_2)
=Z_2;\nonumber \\
\overline{X}_1O(1,0)&=&P_{\theta_1}^{1,2}X_1Z_2P_{\theta_1}^{1,2}X_1
=P_{\theta_1}^{1,2}X_1P_{\theta_1}^{1,2}X_1Z_2=Z_2,
\end{eqnarray}
so that $[\overline{X}_1,O(1,0)]=0$;
\begin{eqnarray}
O(0,1)\overline{X}_1&=&P_{\theta_1}^{1,2}\left(P_{\theta_2}^{2,3}X_2
P_{\theta_1}^{1,2}X_1\right)Z_2\nonumber \\
\overline{X}_1O(0,1)&=&P_{\theta_1}^{1,2}X_1Z_2P_{\theta_1}^{1,2}
P_{\theta_2}^{2,3}X_2
=-P_{\theta_1}^{1,2}\left(X_1P_{\theta_1}^{1,2}P_{\theta_2}^{2,3}X_2\right)Z_2,
\end{eqnarray}
so that one need only compare the terms in parentheses:
\begin{eqnarray}
P_{\theta_2}^{2,3}X_2P_{\theta_1}^{1,2}X_1&=&\left(|0\rangle_2\langle 1|X_3
M_{\theta_2}^3X_3+|1\rangle_2\langle 0|M_{\theta_2}^3\right)\left(
|0\rangle_1\langle 1|X_2M_{\theta_1}^2X_2
+|1\rangle_1\langle 0|M_{\theta_1}^2\right)\nonumber \\
&=&\left(\tan\theta_1|0\rangle_1\langle 1|+\cot\theta_1|1\rangle_1\langle 0|
\right)|0\rangle_2\langle 1|X_3M_{\theta_2}^3X_3
+\left(\cot\theta_1|0\rangle_1\langle 1|+\tan\theta_1|1\rangle_1\langle 0|
\right)|1\rangle_2\langle 0|M_{\theta_2}^3;\nonumber \\
X_1P_{\theta_1}^{1,2}P_{\theta_2}^{2,3}X_2&=&\left(|1\rangle_1\langle 0|
X_2M_{\theta_1}^2X_2+|0\rangle_1\langle 1|M_{\theta_1}^2\right)
\left(|0\rangle_2\langle 1|X_3
M_{\theta_2}^3X_3+|1\rangle_2\langle 0|M_{\theta_2}^3\right)\nonumber \\
&=&\left(\tan\theta_1|0\rangle_1\langle 1|+\cot\theta_1|1\rangle_1\langle 0|
\right)|0\rangle_2\langle 1|X_3M_{\theta_2}^3X_3
+\left(\tan\theta_1|1\rangle_1\langle 0|+\cot\theta_1|0\rangle_1\langle 1|
\right)|1\rangle_2\langle 0|M_{\theta_2}^3,\hphantom{aaa}
\end{eqnarray}
which coincide. Therefore, $\overline{X}_1'=(-1)^{g_2}\overline{X}_1$. The
transformations on $\overline{Z}$ and $\overline{X}$ by the 
$\mathbb{Z}_2\times\mathbb{Z}_2$ operators are equivalent to conjugation under
an effective operator $O_{\rm eff}(g_1,g_2)=\overline{X}^{g_1}
\overline{Z}^{g_2}$, where $\overline{X}$ and $\overline{Z}$ for the left
boundary are defined in Eq.~(\ref{eq:effectivePaulis}. A similar result holds 
for the right boundary. Therefore, $O_{\rm eff}(g_1,g_2)$ has the same form 
as for the regular cluster state, discussed above; this isn't surprising, as
the cluster-state symmetry operators are included in the general form,
Eq.~(\ref{eq:symmetryops}). To summarize: the states given by 
Eq.~(\ref{eq:tensorstate1}) possess $\mathbb{Z}_2\times\mathbb{Z}_2$ SPT order 
for all $\theta_i$, albeit one that is generally neither unitary nor onsite.

\section{Entanglement Spectrum}
\label{app:entanglement}

Consider the entanglement spectrum for a bipartition of the state defined by 
the MPS matrices~(\ref{eq:BUtensorJ}), where the boundary states are defined 
in Eq.~(\ref{eq:tensorproducts}), leading to the state given in 
Eq.~(\ref{eq:tensorstate1}).  Applying Eq.~(\ref{eq:Phi}) yields the rather 
intuitive-looking expressions
\begin{equation}
\Phi_{\ell,j}^{R}=\begin{cases}
x_1|0\rangle & (\ell,j)=(n-1,0);\cr
x_2|1\rangle & (\ell,j)=(n-1,1);\cr
\frac{1}{\sqrt{2}}\left(\prod_{k=\ell+1}^{n-1}
C_{\theta_k}^{k,k+1}\right)
\left(x_1|0\rangle+x_2|1\rangle\right)\otimes|+\rangle^{\otimes(n-\ell-2)}
\otimes|j\rangle & \ell < n-1,
\end{cases}
\label{eq:PhiR}
\end{equation}
and
\begin{equation}
\Phi_{\ell,j}^{L}=\begin{cases}
a_L\cos\theta_1|0\rangle+b_L\sin\theta_1|1\rangle 
& (\ell,j)=(1,0);\cr 
a_L\sin\theta_1|0\rangle-b_L\cos\theta_1|1\rangle & 
(\ell,j)=(1,1);\cr
\frac{1}{\sqrt{2}}\left(\prod_{k=1}^{\ell-1}C_{\theta_k}^{k,k+1}\right)
\begin{pmatrix}
\cos\theta_{\ell} & \sin\theta_{\ell} \cr
\sin\theta_{\ell} & -\cos\theta_{\ell} \cr
\end{pmatrix}|j\rangle\otimes|+\rangle^{\otimes(\ell-2)}\otimes
\left(a_L|0\rangle+b_L|1\rangle\right) & \ell > 1,
\end{cases}
\label{eq:PhiL}
\end{equation}
where $x_1$ and $x_2$ are defined below Eq.~(\ref{eq:tensorstate1}). The 
$\Phi^{L}_{\ell,j}$ and $\Phi^{R}_{\ell,j}$ correspond to 
$2^{\ell}$-dimensional and $2^{n-\ell}$-dimensional vectors respectively, whose 
index gets summed over in Eq.~(\ref{eq:V}). A straightforward calculation 
employing Eqs.~(\ref{eq:PhiR}) and (\ref{eq:PhiL}) yields
\begin{eqnarray}
\langle 0|V_{\ell}^{R}|0\rangle&=&\begin{cases}
|x_1|^2 & \ell=n-1;\cr
\frac{1}{2}\left[1+\left(|x_1|^2-|x_2|^2\right)\prod_{k=\ell+1}^{n-1}
\cos(2\theta_k)\right] & \ell<n-1;\cr
\end{cases}\nonumber \\
\langle 0|V_{\ell}^{R}|1\rangle&=&\langle 1|V_{\ell}^{R}|0\rangle=0;
\nonumber \\
\langle 1|V_{\ell}^{R}|1\rangle&=&\begin{cases}
|x_2|^2 & \ell=n-1;\cr
\frac{1}{2}\left[1-\left(|x_1|^2-|x_2|^2\right)\prod_{k=\ell+1}^{n-1}
\cos(2\theta_k)\right] & \ell<n-1.\cr
\end{cases}
\label{eq:VmR}
\end{eqnarray}
and
\begin{eqnarray}
\langle 0|V_{\ell}^{L}|0\rangle&=&\begin{cases}
|a_L|^2\cos^2\theta_1+|b_L|^2\sin^2\theta_1 & \ell=1;\cr
\frac{1}{2}\left[1+\left(|a_L|^2-|b_L|^2\right)\prod_{k=1}^{\ell}
\cos(2\theta_k)\right] & \ell>1;\cr
\end{cases}\nonumber \\
\langle 0|V_{\ell}^{L}|1\rangle&=&\langle 1|V_{\ell}^{L}|0\rangle
=\begin{cases}
\left(|a_L|^2-|b_L|^2\right)\cos\theta_1\sin\theta_1 & \ell=1;\cr
\frac{1}{2}\left(|a_L|^2-|b_L|^2\right)\sin(2\theta_{\ell})\prod_{k=1}^{\ell-1}
\cos(2\theta_k) & \ell>1;\cr
\end{cases}\nonumber \\
\langle 1|V_{\ell}^{L}|1\rangle&=&\begin{cases}
|a_L|^2\sin^2\theta_1+|b_L|^2\cos^2\theta_1 & \ell=1;\cr
\frac{1}{2}\left[1-\left(|a_L|^2-|b_L|^2\right)\prod_{k=1}^{\ell}
\cos(2\theta_k)\right] & \ell>1.\cr
\end{cases}
\label{eq:VmL}
\end{eqnarray}
Consider the case of a bulk bipartition, $2<\ell<n-1$. One obtains
\begin{equation}
V_{\ell}^{R}=\frac{1}{2}\begin{pmatrix}
1+\alpha & 0\cr
0 & 1-\alpha\cr
\end{pmatrix};\;
V_{\ell}^{L}=\frac{1}{2}\begin{pmatrix}
1+\beta\cos(2\theta_{\ell}) & \beta\sin(2\theta_{\ell})\cr
\beta\sin(2\theta_{\ell}) & 1-\beta\cos(2\theta_{\ell})\cr
\end{pmatrix},
\label{eq:VLVR}
\end{equation}
where
$\alpha:=\left(|x_1|^2-|x_2|^2\right)\prod_{k=\ell+1}^{n-1}\cos(2\theta_k)$ and 
$\beta:=\left(|a_L|^2-|b_L|^2\right)\prod_{k=1}^{\ell-1}\cos(2\theta_k)$. The 
eigenvalues of $V_{\ell}^{R}V_{\ell}^{L}$ are readily obtained:
\begin{equation}
\lambda_{\pm}=\frac{1}{4}\left\{1+\alpha\beta\cos(2\theta_{\ell})
\pm\sqrt{[1+\alpha\beta\cos(2\theta_{\ell})]^2-(1-\alpha^2)(1-\beta^2)}
\right\}.
\label{eq:spectrumapp}
\end{equation}
Eq.~(\ref{eq:spectrumapp}) corresponds to the entanglement spectrum of the 
(unnormalized) state defined by Eqs.~(\ref{eq:BUtensorJ}) and 
(\ref{eq:tensorstate1}).

\end{widetext}

\bibliography{ref}

\end{document}